\documentclass[twocolumn]{aastex631}

\shorttitle{AutoWISP}
\shortauthors{Romero et al.}
\graphicspath{{./}{figures/}}

\usepackage{import}
\usepackage{relsize}
\usepackage{amsmath}

\graphicspath{ {figures/} }

\begin{document}
\title{AutoWISP \footnote{\url{https://github.com/kpenev/AutoWISP}}: Automated Processing of Wide-Field Color Images}

\author[0000-0001-9436-6027]{Angel E. Romero}
\affiliation{
    University of Texas at Dallas,
    800 W. Campbell Road,
    Richardson, TX 75080-3021, USA
}
\email{angel.romero@utdallas.edu}

\author[0000-0003-4464-1371]{Kaloyan Penev}
\affiliation{
    University of Texas at Dallas,
    800 W. Campbell Road,
    Richardson, TX 75080-3021, USA
}
\email{kaloyan.penev@utdallas.edu}

\author[0000-0002-9490-2093]{S. Javad Jafarzadeh}
\affiliation{
    University of Texas at Dallas,
    800 W. Campbell Road,
    Richardson, TX 75080-3021, USA
}
\email{ashkan.jafarzadeh@utdallas.edu}

\author[0000-0002-8423-0510]{Zoltan Csubry}
\affiliation{
    Department of Astrophysical Sciences, Princeton University, NJ 08544, USA
}
\email{zcsubry@astro.princeton.edu}

\author[0000-0001-8732-6166]{Joel D. Hartman}
\affiliation{
    Department of Astrophysical Sciences, Princeton University, NJ 08544, USA
}
\email{jhartman@astro.princeton.edu}

\author[0000-0001-7204-6727]{G{\'a}sp{\'a}r {'A}. Bakos}
\affiliation{
    Department of Astrophysical Sciences, Princeton University, NJ 08544, USA
}
\email{gbakos@astro.princeton.edu}



%
\begin{abstract}
We have developed a software pipeline, AutoWISP, for extracting high-precision photometry from citizen scientists' observations made with consumer-grade color digital cameras (digital single-lens reflex, or DSLR, cameras), based on our previously developed tool, AstroWISP. The new pipeline is designed to convert these observations, including color images, into high-precision light curves of stars. We outline the individual steps of the pipeline and present a case study using a Sony-$\alpha$ 7R II DSLR camera, demonstrating sub-percent photometric precision, and highlighting the benefits of three-color photometry of stars.  Project PANOPTES will adopt this photometric pipeline and, we hope, be used by citizen scientists worldwide. Our aim is for AutoWISP to pave the way for potentially transformative contributions from citizen scientists with access to observing equipment.

\end{abstract}
%
%
\section{Introduction}
\label{c:intro}
Photometry is the direct measurement of light from stars and is of fundamental importance to astronomy. Through photometry, astronomers can detect exoplanets via transit observations, determine a star's distance and size, measure stellar rotation using star spots, characterize supernovae, and gain valuable insights into a star's structure and characteristics, among other applications. Photometry is one of the most essential observational techniques in astronomy \citep{Henden_82}. 

In recent years, the time and effort dedicated by professional astronomers to photometry have grown drastically. This is evident in the numerous ground-based exoplanet surveys (e.g., HATNet \citep{Bakos_et_al_04}, HATSouth \citep{Bakos_et_al_13}, Super-WASP \citep{Pollaco_et_al_06, Hellier_et_al_11}, KELT \citep{Pepper_et_al_07}, and MEarth \citep{Nutzman_Charbonneau_08}), microlensing surveys (e.g., KMTNet \citep{Kim_et_al_10}, MOA \citep{Bond_et_al_01}, MicroFUN, and OGLE \citep{Udalski_et_al_02}), whole-sky surveys (e.g., LSST, Evryscope \citep{Law_et_al_15}, and HATPI \citep{HATPI_18}), and space missions (e.g., Kepler \citep{Koch_2010}, K2 \citep{Howell_14}, TESS \citep{Ricker_et_al_09}, and PLATO \citep{rauer2024platomission}), many of which have observed and measured hundreds of thousands of stars, collectively detecting thousands of exoplanets and significantly advancing their detection.

One of the most common methods for detecting exoplanets is transit detection. Transit detection looks for dips in starlight caused by planets passing in front of stars (i.e., dips in light curves). In contrast, radial velocity detection uses spectroscopy to identify changes in the star's radial velocity as it orbits around the barycenter of the planetary system. Most ground-based surveys have set up arrays of CCD detectors and use the transit method to detect exoplanets. However, the CCDs used in these surveys typically cost at least 10,000 USD, such as the Apogee Alta U16M CCD camera, which was used in both HATSouth and KELT South \citep{Bakos_et_al_13, Pepper_et_al_12}. This creates a significant barrier to the creation and expansion of exoplanet surveys, which we hope to address through the use of inexpensive detectors and the efforts of citizen scientists.

\subsection{Motivation}
\label{s:goals}
There is a need for photometric follow-up to the large population of interesting objects found by these previously mentioned surveys and space missions. However, professional astronomers cannot keep up with these efforts due to the overwhelming number of targets that can benefit from follow-up. Many of these objects only require modest telescopes for high-quality photometric characterization but are limited to using sophisticated and expensive detectors (CCDs). If precision photometry is made possible using inexpensive detectors, such as high-quality off-the-shelf digital cameras (digital single-lens reflex (DSLR) cameras), with user-friendly software, then enthusiastic citizen scientists can undertake many observations in support of professional astronomy; citizen scientists are already a prominent group in the astronomical community, and utilizing their abilities has the potential to lead to discoveries and assist professionals in astronomy. The highly respected American Association of Variable Star Observers has promoted citizen scientist photometry for over a century. In the area of exoplanets, several projects are currently working to harness the power of citizen scientist photometric observations, such as: the JPL-lead Exoplanet Watch (EW) \citep{Zellem_et_al_19}, project PANOPTES \citep{Guyon_et_al_14}, and the Unistellar network \citep{Esposito_et_al_21}, to name a few. These and numerous other examples of citizen scientists providing observation data demonstrate that amateur astronomers are eager to contribute their skills, time, and equipment to push science forward. Some examples of the kinds of observations that citizen scientist can contribute to include but are not limited to: identifying false positives among transiting planet candidates \citep{Zellem_et_al_19, Zellem_et_al_20}, photometric followup of known exoplanets, stellar variability and rotation \citep{AAVSO_14}(https://www.aavso.org/), solar system objects \citep{TargetAsteroids_18}, galactic novae and bright supernovae.

Furthermore, enabling high-precision photometry to be extracted from observations with inexpensive, easy-to-use detectors, such as off-the-shelf consumer DSLR cameras, can dramatically expand the pool of citizen scientists capable of participating in photometric follow-up efforts. However, DSLRs have significant disadvantages when performing photometry, such as lower quantum efficiency and smaller well capacity (which increases Poisson noise and limits photometric precision), higher dark currents at increased temperatures (which create more noise in the detector), and the presence of a Bayer mask superimposed on their sensors (which alters how light falls onto a pixel of a given color) \citep{Zhang_et_al_15}. Despite these drawbacks, DSLRs remain very inexpensive to obtain, set up, and operate for photometry. Additionally, the Bayer masks of DSLRs allow for simultaneous three-color photometry, which is difficult and expensive to achieve with CCDs.


Currently, there are a plethora of photometric solutions available to the professional astronomy community (e.g. "Photutils" \citep{Bradley_et_al_17}; “AstroImageJ” \citep{Collins_et_al_16}; “ISIS” \citep{Alard_Lupton_98}; “FITSH” \citep{Pal_12}; and “ATP” \citep{Laher_et_al_12}, listing the most well known), but only very few are used or accessible to citizen scientists (e.g. AIP4WIN \citep{AIP4WIN_18}, MaximDL \citep{MaximDL_18}, Aperture Photometry Tool (APT) \citep{APT_18}, IRIS \citep{IRIS_18}, and AstroImageJ \citep{Collins_et_al_16} being the most commonly used). However, to date, there has been no milli-magnitude precision photometry from wide fields of view citizen scientist images. Furthermore, none of the tools used by citizen scientists simultaneously handle Bayer masks, work across operating systems, and allow automating the processing of hundreds or thousands of images. To ensure further accessibility, we are creating a browser user interface (BUI) that will provide an intuitive interface for setting up the processing, tracking its progress, and reviewing results. However, the BUI is outside the scope of this paper and will be described in future publications by our group.

In this paper, we discuss the creation of an automated pipeline, AutoWISP, which uses our photometric tools AstroWISP \footnote{{\url{https://github.com/kpenev/AstroWISP}}} \citep{Penev_et_al_25}. Section \ref{c:methodology}, details the methodology of AutoWISP. Section \ref{c:implementation} the implementation of AutoWISP. Section \ref{c:results} is a case study of our pipeline, demonstrating the capabilities of AutoWISP when applied to observations obtained with a Sony-$\alpha$7R DSLR Camera. Finally, Section \ref{c:discussion} describes the future of AutoWISP.
\section{Methodology}
\label{c:methodology}
The pipeline consists of 6 major steps: image calibration, astrometry, photometry, magnitude fitting, creating light curves, and light curve post-processing. We present the entire collection of processing steps and scripts so citizen scientists can use them.
\begin{figure}
    \centering
    \includegraphics[width=0.5\linewidth]{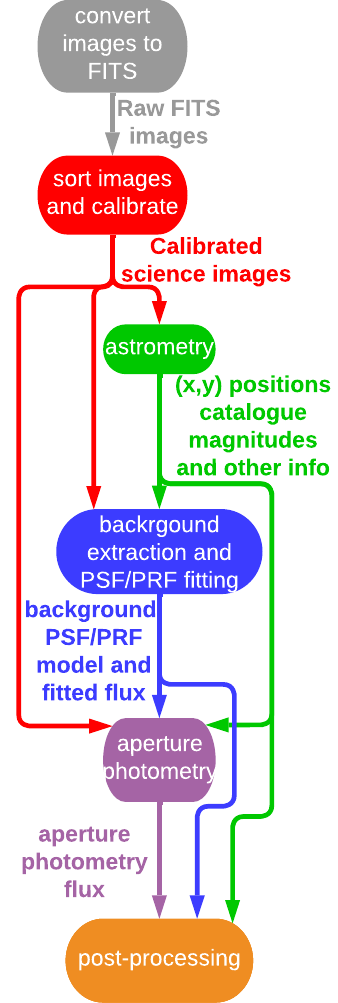}
    \caption{The image processing pipeline steps and their products. The arrows indicate the products of each step and where they will be used.}
    \label{fig:pipeline}
\end{figure}
Our general photometry procedure can be visually seen in Figure \ref{fig:pipeline} and will be explained in further detail in Sections \ref{s:calibration}-\ref{s:post_processing}.

In brief, however, the general procedures are as follows:
\begin{enumerate}
    \item Users input their raw fits images (flat, dark, bias, or object image types as specified from their fits headers). These images are then pre-processed, where master calibration frames are generated, then frames are calibrated with these masters. This is a similar implementation as used by the HATNet, HATSouth, and HATPI projects \citep{Bakos_et_al_04, Bakos_et_al_13, HATPI_18}
    \item After the pre-processing, we extract the source (star) positions from the images and perform astrometry (plate-solving) to find a transformation that allows us to map sky coordinates (RA, Dec) into image coordinates. This allows the use of external catalogue data for more precise positions of the sources than can be extracted from survey images and the use of auxiliary data provided in the catalogue about each source in the subsequent processing steps of the pipeline.
    \item Next, for each calibrated object frame, we extract flux measurements, background level, and uncertainties for the catalogue sources found in the image, which map to some position within the frame using the astrometric transformation derived in the previous step. This step is performed using AstroWISP \citep{Penev_et_al_25}, and we refer the reader to that article for a detailed description. We briefly summarize the process here for completeness. 
    
    AutoWISP takes into account the response of the pixels due to the fact that the same amount of light falling on one part of the pixel is likely to produce a different response in a different part of the pixel, this is what we call sub-pixel sensitivity. Non-uniform sensitivity on a sub-pixel level affects the PRF in an image, and the effect depends on where within a pixel the center of the source lies. Thus, it is necessary to correct for the sub-pixel sensitivity variations when deriving PSFs for a given image. For example, for DSLR or color images, we can consider the Bayer mask, which is a filter that is superimposed on the detector with arrangement of pixels sensitive to different colors in super-pixels (see Figure \ref{fig:bayer_mask}). Since each color accounts for 1/4th of the super-pixel, the Bayer mask is an extreme version of the varying amount of the sub-pixel sensitivity. When processing a singular color channel, 3/4th of the pixel area can be considered completely insensitive to light.
    
    There are many flavors of photometry. This pipeline supports: point spread function (PSF) or pixel response function (PRF) fitting (where the PRF is the PSF convolved with the sub-pixel sensitivity), and aperture photometry, with aperture photometry requiring PSF fitting.
        \begin{enumerate}
            \item For each point source, we first measure the distribution of light on the detector (the PSF). The idea of PSF fitting is to model that distribution as some smooth parametric function centered on the projected source position with an integral equal to 1. The flux of the source is then found from a least squares scaling between the predicted and observed pixel values. While the flux will differ for each star, the shape parameters are assumed to vary smoothly as a function of source properties and position within the image. In principle, smooth changes can also be imposed across images, though that requires a highly stable observing platform in practice. For our implementation, we also take the sub-pixel sensitivity into account due to the non-uniform sensitivity of the sub-pixel level as previously described. Lastly, we store the PSF information for later use during aperture photometry. 
            \item Similar to the PSF, we can perform PRF fitting. The PRF can be thought of as a super-resolution image of the light of a star falling on the individual pixels, where it is represented as a continuous piecewise polynomial function of sub-pixel position on each pixel (i.e., the PSF convolved with the sub-pixel sensitivity).
            \item After PSF fitting, we perform aperture photometry, a photometry method that sums the flux within a circular aperture centered on each source. For aperture photometry, we correct for non–uniform pixels by using the sub-pixel sensitivity information/map and adequately integrate the PSF model to determine the fractional contributions of pixels straddling the aperture boundary.
        \end{enumerate}
    \item After extracting flux measurements, we perform ensemble magnitude fitting. The photometry of individual frames is calibrated to the photometry of a reference frame by applying a correction as a smooth function of image position, brightness, color, and other user-specified parameters. This procedure excludes stars showing significantly larger variability than other similarly bright stars and is repeated multiple times, where the reference frame is replaced with a stack of many frames corrected in the previous iteration.
    \item Next, we create light curves. This is a transpose operation, collecting the photometry of each star from all images and putting them in a single file (the light curve)
    \item Finally, after producing light curves, we perform post-processing using external parameter decorrelation (EPD) and trend filtering algorithm (TFA) to correct effects not corrected during magnitude fitting, further improving photometric precision. These procedures are explained in greater detail in the following articles \citep{Huang_et_al_15b, Bakos_et_al_13, Zhang_et_al_15, Kovacs_Bakos_Noyes_05}. 
\end{enumerate}

The following sections provide the exact details of the automation besides the PSF/PRF fitting and aperture photometry, since those are explained in our AstroWISP article \citep{Penev_et_al_25}.

\subsection{Calibration:}
\label{s:calibration}

After users input their raw fits images, as classified by their corresponding headers, the first step is to individually calibrate each image and then combine all the calibrated images into master frames (master biases, master darks, master flats), following algorithms used by the HATSouth project \citep{Bakos_et_al_13}. These masters are then applied to the OBJECT frames to obtain calibrated versions (i.e. calibrated object frames) used by all subsequent steps. The sequence of steps is as follows:

\begin{enumerate}
    \item Calibrate raw bias frames.
    \item Generate master bias frames.
    \item Calibrate raw dark frames using the master biases.
    \item Generate master dark frames.
    \item Calibrate raw flat frames using the master biases and master darks.
    \item Generate master flat frames.
    \item Calibrate raw object frames using all masters.
\end{enumerate}

The overall calibration for object frames is given by

\begin{equation}
    C(I)= \frac{I - O(I) - B_0 - (\frac{\tau[I]}{\tau[D_0]})D_0}{F_0/||F_0||}
\end{equation}

Where $I$, $O(I)$, and $C(I)$ represent the image, over-scan region, and calibrated image respectively; $B_0$, $D_0$, and $F_0$ represents masters calibration images of bias, dark, and flat respectively; $\tau[I]$ and $\tau[D_0]$ are the exposure times of the image and dark frame respectively \citep{Pal_09}. These master frames are stacks of individually calibrated bias, dark, and flat frames. As a result, their signal-to-noise ratio is significantly increased compared to individual unstacked frames, allowing for much better calibration. In each case, the frames are split into groups where the measured effect is not expected to vary. For example a separate master can be generated for each night of observing, or dark frames can be split into groups with similar temperature etc.. The individual frames are stacked, with outlier pixels being discarded. Additionally, when master flats are generated, flat frames containing clouds or with low signal-to-noise are discarded. These final calibrated images consist of three extended fits images, which include the pixel values, error estimates, and the integer mask with flags indicating saturation, leak, etc.

Two noteworthy features are: 
\begin{enumerate}
    \item The error is rigorously tracked for each individual pixel of each image through each step. This allows us to produce variance images used during photometry extraction for accurate error estimates, which typically improves the photometric precision. 
    \item The procedure for generating master flats is inherited from the HAT projects \citep{Bakos_et_al_13}, allowing images of the dusk sky to be used for robust flat-field corrections. Particular noteworthy features include automatic cloud detection, optimal treatment of variable sky brightness on large spatial scales, and safeguards against bright stars showing up above the sky-noise level.
\end{enumerate}

\subsubsection{Mask creation}
Before calibrating, we first create a mask image to exclude from further processing. This includes pixels identified as saturated pixels, or after additional masking is done from master generation. Saturated pixels are flagged if they exceed a specified value, close to the maximum value the pixel can have, indicating that either the pixel has accumulated close to the maximum number of electrons it can hold or the largest amplifier value has been reached. Also, if they are adjacent to saturated pixels in the leak direction where they could receive a leaked charge, they are flagged as saturated pixels. Additional masks (supplied by the user) can be applied to flag hot pixels, non-linear pixels etc, which will then be properly propagated to the photometry measurements. Bad pixels can be flagged by the users, identifying where hot pixels (pixels with mean dark current significantly larger than the dark current of normal pixels), pixels which exhibit high non-linear readout, pixels with large readout noise, or dead pixels (pixels which don't accumulate any charge) are located on the detector. Furthermore, we transfer any masks in the masters used, including taking separate mask-only FITS files.

\subsubsection{Overscan corrections}
Afterwards, we apply "overscan" corrections. In many instances, the imaging device provides extra areas that attempt to measure bias level and dark current, e.g. by continuing to read pixels past the physical number of pixels in the device, thus measuring the bias or by having an area of pixels which are somehow shielded from light, thus measuring the dark level in real time. Such corrections are superior to the master frames in that they measure the instantaneous bias and dark level, which may vary over time due to, for example, the temperature of the detector varying. However, bias level and dark current, in particular, can vary from pixel to pixel, which is not captured by these real-time areas. Hence, the best strategy combines both, which differs for different detectors. The pipeline allows (but does not require) such areas to be used to estimate some smooth function of image position to subtract from each raw image, and then the masters are applied to the result. This works mathematically because the masters will also have their values corrected for the bias and dark measured by these areas from the individual frames used to construct them. In this scheme, the master frames are used only to capture the pixel-to-pixel differences in bias and dark current. We refer to these areas as "overscan", although that term means only one type of such area, where the unifying theme is that the "overscan" areas do not recieve light.

\subsubsection{Subtraction of bias level and dark current}
While, and if, overscan corrections are applied to all the raw frames, the first round of subtractions of the bias level and dark current is done to each frame (i.e., de-biasing and de-darking). This is done by subtracting the master bias and the master dark from the target image. However, the master bias is not subtracted from the raw bias frames since calibrating those generates the master bias. Similarly, for the raw dark frames, master bias corrections are applied, but master darks are not, and for the raw flat frames, master dark and master bias corrections are applied, but master flats are not. All other image types get the complete set of corrections. Master bias and master dark images have pixels calculated as the median with outlier rejection of the corresponding pixels in individual bias and dark images, respectively. 

\subsubsection{Flat field corrections are applied}
Next, we apply a flat field correction to all the object frames, skipping individual raw bias, dark, and flat frames. This is done by taking the ratio of the bias and dark corrected frames and the master flat, pixel by pixel. However, properly generating a master flat involves much more than the prior two. Since flat frames can be images of the sky, special care must be taken to compensate for changes in the large-scale structure from one calibrated flat frame to the other. Furthermore, with sky frames, there is always the possibility of clouds, so we have an automated procedure for detecting clouds in individual frames, discarding them, or detecting cloudy flat collections and refusing to generate a master flat altogether. The procedure follows similarly to HATSouth \citep{Bakos_et_al_13}.

\subsubsection{Individual pixel errors are calculated}
The de-biasing and de-darking adds the (slight) noise in the master frames (and the overscan corrections), and the scaling by the flat introduces the error in the master flat but also changes the "gain" differently for each pixel \citep{Newberry_91}. To properly handle all those, calibrating raw frames in the pipeline produces two images: the calibrated image and an error image, taking into account the gain, containing the estimated standard deviation in the calibrated image pixel values.

\subsection{Source Extraction}
\label{s:source_extraction}
We have an external tool, fistar (part of the FITSH package \citep{Pal_12}), that extracts sources from the images. Fistar finds and characterizes point-like sources, providing (x, y) centroid coordinates, an estimate of the source's flux, an elliptical Gaussian fit for the source's shape, and a signal-to-noise estimate. Filtering is done based on the flux estimates and shape parameters to remove outliers. This process is packaged in our existing software suite, AstroWISP \citep{Penev_et_al_25}, for all operating systems, without requiring compilation or complicated installation steps. 

\subsection{Astrometry Based Photometry:} 
\label{s:astrometry}
Before extracting photometry, we must know the stars' locations in the calibrated images. Locations can be obtained using the source extraction in Section \ref{s:source_extraction}. However, for wide-field images, better positions can be obtained by deriving a transformation between an existing much higher resolution catalogue (e.g., Gaia (\citealp{GAIA_2016, GAIA_2023})) and the stars extracted from the image (astrometry) and using it to project the catalogue positions on the frame. This results in more precise and accurate positions because the transformation is a smooth function with much fewer parameters than the number of stars used to derive it \citep{Bakos_et_al_04}. Tools that derive astrometric solutions already exist, most notably Astrometry.net \citep{Lang_et_al_10}. However, Astrometry.net uses only a few stars for its solutions, making it unusable for high-order transformations. Since high-order transformations are required for wide-field images (e.g., the HAT surveys or DSLR data) with thousands of stars in each image, we provide our own tool.

\begin{figure}
    \centering
    \includegraphics[width=0.45\textwidth]{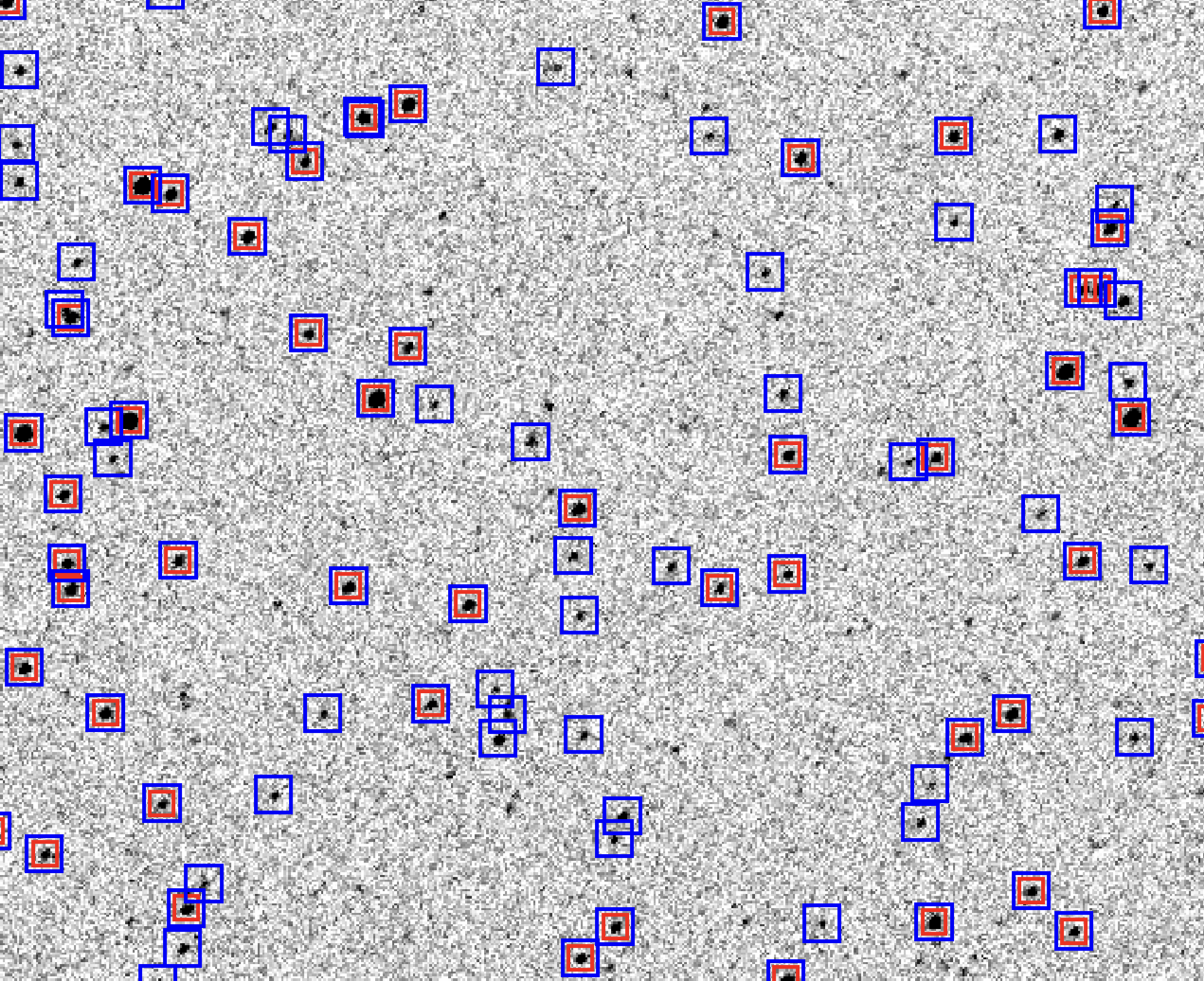}
    \caption{The source extraction versus catalogue projections of our astrometry step placed on top of the corresponding FITS image, where blue squares are the catalogues projected sources and red squares are the extracted sources from our astrometry}
    \label{fig:Astrometry}
\end{figure}

We begin by obtaining an Astrometry.net solution, which provides an initial approximation for the transformation from sky to image coordinates. This transformation is then used to project a list of Gaia stars to image coordinates. We assume that if a catalogue source projects within a user-specified tolerance (e.g., one pixel) of an extracted source, it must be the same star, and we use this match to derive a new transformation. The process is then repeated with the updated transformation. This cycle continues until the change in projected positions from one step to the next falls below a threshold (e.g., 0.005 pixels). The solution succeeds if more than 90\% of the extracted sources are matched to a catalogue source. This can be visually seen in Figure \ref{fig:Astrometry}. We note in Figure \ref{fig:Astrometry} the catalogue is queried deeper than the limit imposed on source extraction.

We currently have an implementation based on Gaia catalogues \citep{GAIA_2016, GAIA_2023}, which has high-precision information on stars' positions, proper motion, and magnitude, which allows us to use up-to-date stellar positions and have much better information about blended sources, binary stars, source properties, etc.. This information is then used during photometry extraction. Given Gaia's unprecedented astrometric precision, the accuracy and precision of source positions in this step should be well beyond what is necessary for high-precision photometry, contributing only negligibly to the photometric error. 

Once we have the transformation, we use it to project the sky coordinates (RA, Dec) from the catalogue to image coordinates (x, y). These (x, y) positions are then what we call the center of the source in PSF fitting (described in AstroWISP \citep{Penev_et_al_25}) and/or the position that is used to define the background annulus and exclude pixels near sources.

\subsection{Photometry:} 
\label{s:photometry}
The prescription for doing point spread function (PSF), or pixel response function (PRF), photometry and aperture photometry is precisely what we describe in our AstroWISP paper \citep{Penev_et_al_25}, as briefly described in the summary of each step in Section \ref{c:methodology}. The entire pipeline is set up to use magnitudes instead of fluxes, which are converted from relative fluxes using the formula $Mag = Z -2.5 \cdot \log_{10}(Flux/t_{exp})$, where Z is the magnitude that corresponds to a flux of 1 ADU/s and $t_{exp}$ is the exposure time in seconds.

\subsection{Magnitude Fitting:}
\label{s:mag_fit}
In ground-based applications, the night sky is imaged through a variable amount of atmosphere, which is subject to changes (i.e., clouds, humidity, etc.). In addition, various instrumental effects are generally present (e.g., detector sensitivity, gain or dark current drifts due to temperature fluctuations, detectors having physical wires on pixels to prevent charge leakage on neighboring pixels). Most photometry projects apply empirically determined corrections \citep[e.g.][]{Pal_09}. We apply these in order of increasing risk or probability of suppressing real astrophysical variability, and the photometry at any stage of post-processing (including completely unprocessed) will be preserved in the final light curves. The magnitude fitting step aims to eliminate as many possible effects that modify the measured source brightness within an image in a manner that depends smoothly on the properties of the source or its location in the image.

In the pipeline, this is done several times (iteratively). The first time, a single frame which appears to be of very high quality (sharp PSF, high atmospheric transparency/low airmass, dark sky, and/or small astrometric residuals) is used as the reference frame. The corrected brightness measurements of the individual frames are stacked to produce a much higher signal to noise "master reference frame", which is then used in a second iteration of the magnitude fitting process, the procedure is repeated until the reference stops changing. We currently do this for both PSF photometry and aperture photometry magnitudes. PSF fitting will work better for some stars, and for others, aperture photometry will work better. Furthermore, magnitude fitting is done separately for each color channel. 

The procedure is as follows:

Each frame is compared to the reference frame. We compute the difference for every star $i$, $\Delta m = m^{i} - {m_{ref}}^{i}$, where $m^i$ and ${m_{ref}}^i$ represent the raw magnitudes in the target and reference frames, respectively. We then use a weighted fit for these differences using a general combination of nth-order polynomials. These polynomials can depend on several factors, including (but not limited to) star position ($x_i$, $y_i$), detrending correction parameters (e.g., catalogue magnitude, color indices of the star), and background level. The weights are determined by the formal errors derived during the photometry step(s).

\subsection{Light Curves:}
\label{s:light_curves}
From the previous steps, photometry is extracted simultaneously for all sources in a given image. To study each source's variability, the measurements from all frames for that source must be collected together. This step performs that reorganization of the data. Each catalogue source's available measurements from the individual frames are collected in a single file, possibly combined with earlier measurements from a different but overlapping telescope pointing or with another instrumental set-up.

\subsection{Post Processing:} 
\label{s:post_processing}
After light curves are generated, we perform two post-processing steps, External Parameter Decorrelation (EPD) \citep{Bakos_et_al_10} and Trend-Filtering Algorithm (TFA) \citep{Kovacs_Bakos_Noyes_05}, to reduce any other sources of noise magnitude fitting was not able to deal with.

\subsubsection{External Parameter Decorrelation (EPD)}
EPD aims to remove correlations between magnitudes and external parameters, e.g., temperature, airmass, humidity, subpixel position, stellar profile parameters characterizing the PSF width (called S), and its elongation (D and K). This removes from each light curve the best fit polynomial of user-specified instrumental and other time variable parameters that explain the most variance. Care must be taken when selecting the parameters to de-correlate against, lest they vary on similar timescales as the target signal. If this happens, this step may distort the target signal.

\subsubsection{Trend Filtering Algorithm (TFA)}
The last step in our automation is removing trends common to many stars in the observed field using our implementation of the Trend-Filtering Algorithm (TFA) \citep{Kovacs_Bakos_Noyes_05}. This procedure works well for a wide array of applications, including spaced-based data from the K2 missions (c.f. \citealp{Huang_et_al_15b}), the HAT surveys \citep{Bakos_et_al_13, Bakos_et_al_04}, and DSLR photometry \citep{Zhang_et_al_15}. The idea is that most instrumental effects will similarly affect multiple sources, and thus, signals common to several sources are suspected of being instrumental rather than real astrophysical variability. TFA removes these similar systematic effects across many stars by selecting a representative set of stars, template stars, for all possible systematics. Then, from each star, the linear least squares fit of the template light curves and that star's light curve are subtracted. During TFA, to avoid distorting or removing a signal of known shape, the signal is included in the list of templates and fits simultaneously with the other templates. However, if the amplitude is already known, we do not fit the amplitude and instead pre-subtract the signal and then perform TFA \citep{Kovacs_Bakos_Noyes_05}. Again, this step can potentially distort or eliminate target signals, so it should be used carefully.

\section{Implementation of AutoWISP}
\label{c:implementation}
All tools in AutoWISP were developed in Python 3 with a command line interface (i.e., argparse \footnote{\url{https://docs.python.org/3/library/argparse.html}}, configparse \footnote{\url{https://docs.python.org/3/library/configparser.html}}) with a Python wrapper around AstroWISP. All code is tracked and managed via GitHub and is available on all three major operating systems (Windows, OS X, Linux). This software is publicly available through the python package index \footnote{\url{https://pypi.org/project/autowisp/}}, and the source code is available through GitHub \footnote{\url{https://github.com/kpenev/AutoWISP}} and Zenodo \footnote{\url{https://doi.org/10.5281/zenodo.16135207}}. The software package will eventually include a browser-user-interface and fully automation and tracking through a database. However, these features are planned for the future and are outside the scope of this article.

We have created Python scripts that perform each of the steps introduced in Section \ref{c:methodology}. Users can run each script to run each step via a command-line interface using a configuration file and arguments. This configuration file is structured into sections, each containing key-value pairs that correspond to the command-line arguments. All calibrated images produced by the pipeline are stored internally in FITS format \citep{FITS_81}. For each frame, non-image byproducts (e.g. extracted sources and their properties, astrometric solutions, photometric measurements, etc.) are stored in an HDF5 \citep{HDF5_18} file, which we call a data reduction file. The HDF5 format is extremely flexible, supports several compression algorithms, and has a fully developed python interface (H5py \footnote{\url{https://docs.h5py.org/en/stable/}}). HDF5 library and python interface are both available under BSD-style Open Source licenses. These resulting files are very convenient, compact, and fully self-contained. They have the ability to store all information on how the contents were generated internally, ensuring full reproducibility without reference to the future pipeline or database. For each step, users provide the required input file types: FITS images, HDF5 data reduction files, or HDF5 light curve files. The image processing pipeline steps and their products are shown in Figure \ref{fig:pipeline}, with the arrows indicating the products of each step and where they will be used. The Gaia catalogues that are used throughout the processing are asynchronously queried using the Astroquery Gaia TAP service\footnote{\url{https://astroquery.readthedocs.io/en/latest/gaia/gaia.html}}, so users do not need to have a local copy of the catalogue.

The pipeline accumulates light curves, with a single file for each catalogue object (refer to Section \ref{s:astrometry}) with possibly multiple sets of observations contributing brightness measurements over time. We also use HDF5 file formats for light curves. Again, with these files being fully self-contained, documenting all previous processing steps, including the ones already reported in the data reduction files, with possibly multiple versions of the same measurements residing side-by-side for cases in which several different configurations were used when applying the pipeline tools.


%
\section{Example Application of AutoWISP}
\label{c:results}
\subsection{Observations}
\label{s:observations}
Observations were carried out by attaching a Sony-$\alpha$7R II (ILCE-7RM2) DSLR Camera with a Canon 135mm f/2 lens to HAT10, a telescope unit from the HAT Network of Telescopes (HATNet), located at the Fred Lawrence Whipple Observatory (FLWO) of the Smithsonian Astrophysical Observatory (SAO) \citep{Bakos_2002}. A similar experiment was conducted by \cite{Zhang_et_al_15} with the same HATNet unit but with a Canon EOS 60D DSLR camera and the same Canon 135 f/2 lens. We point the readers to their paper \citep{Zhang_et_al_15} for more information regarding the unit's setup and the software used to take the observations.

\begin{figure}
    \centering
    \includegraphics[width=0.5\linewidth]{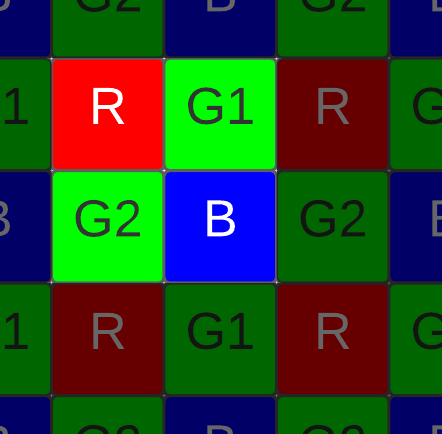}
    \caption{Arrangement of pixels sensitive to different colors in super-pixels (highlighted area) in a DSLR detector}
    \label{fig:bayer_mask}
\end{figure}

The Sony-$\alpha$7R II (ILCE-7RM2) DSLR camera took 8000$\times$5320 pixel (42 megapixels) images using an Exmor R CMOS sensor with a Bayer filter on top of it, with each pixel having a physical size of 4.5$\mu$$\times$4.5$\mu$, readout using a 14-bit amplifier. The Canon 135mm f/2 lens provided a field of view of 7.628$^\circ$. The Bayer mask is laid out in a grid, as seen in Figure \ref{fig:bayer_mask}. Each image of a single color channel only comes from 1/4 of the image area (Bayer masks are organized in 4 channels: red, blue, and two green ones). We distinguish the different green channels of the Bayer mask as Green 1 and Green 2.

\begin{figure*}
    \centering
    \includegraphics[width=0.45\textwidth]{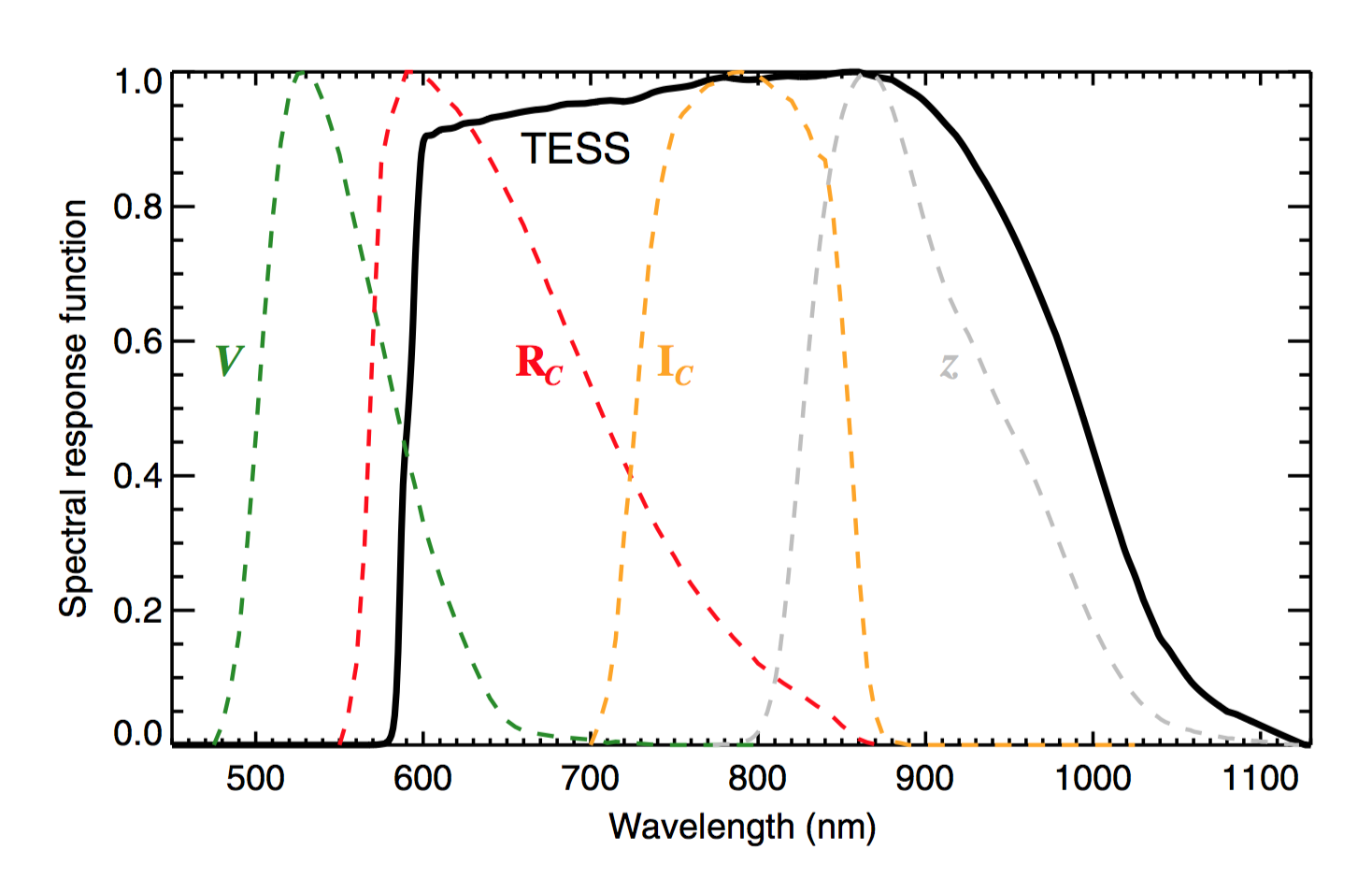}{(a)}
    \includegraphics[width=0.36\textwidth]{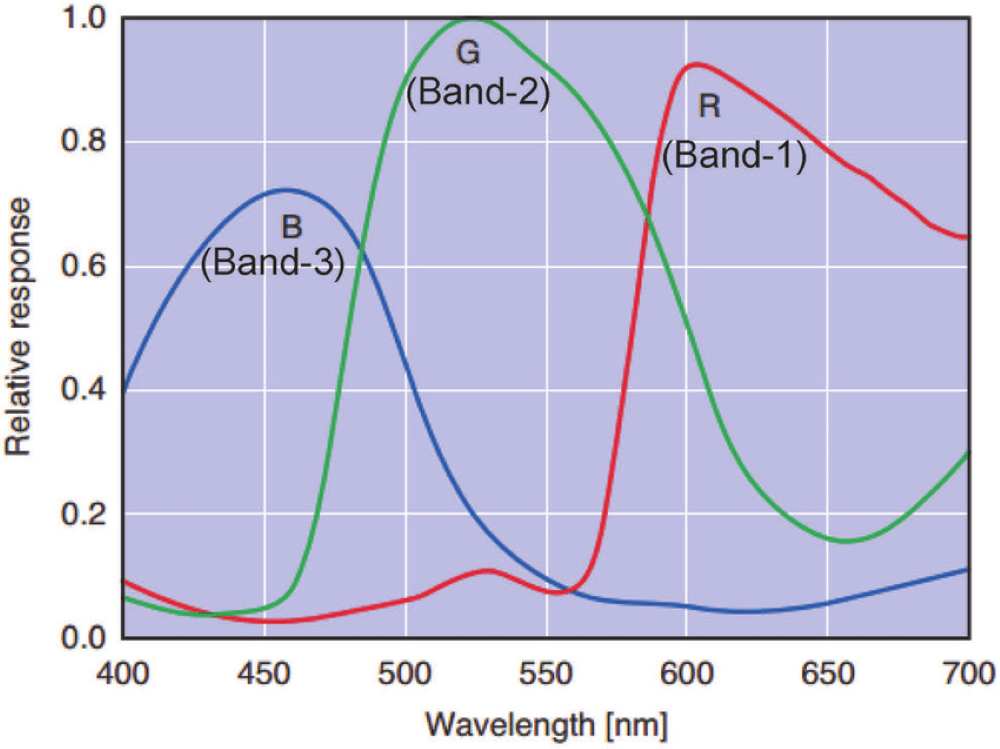}{(b)}
    \caption{The different bandpasses used by (a.) TESS (image credit: \citep{Ricker_et_al_15}) and (b.) the Sony-$\alpha$7R II (ILCE-7RM2) DSLR camera (image credit: \citep{Kiyohisa_16}) (a.) Here, we show the "TESS spectral response function (black line), defined as the product of the long-pass filter transmission curve and the detector quantum efficiency curve. Also plotted, for comparison, are the Johnson-Cousins V, $R_C$, and $I_C$ filter curves and the Sloan Digital Sky Survey z filter curve. Each of the functions has been scaled to have a maximum value of unity" \citep{Ricker_et_al_15}. (b.) The Sony-$\alpha$7R II uses an Exmor R CMOS sensor, and here its spectral response function for each image band is shown, where each function has been scaled to a maximum value of unity as well.}
    \label{fig:bandpass}
\end{figure*}

For comparison, we include the bandpasses used by TESS and the Sony-$\alpha$7R II (ILCE-7RM2) DSLR camera in Figure \ref{fig:bandpass}, which is used in one of the analyses of Section \ref{s:example_lcs}. In this figure, the TESS bandpass spans from the visible red into the infrared wavelengths, whereas the bandpasses of the Sony-$\alpha$ CMOS sensor are confined to the visible spectrum. In order of increasing wavelength, the bandpasses are: Sony-$\alpha$ blue, green, and red, followed by TESS, which primarily covers the infrared range.

%
%


Observations started on 2017-01-11 and ended on 2017-05-28. There are 21724 images, most of which were taken at 30-second exposures. Table \ref{tab:imagetype} shows the breakdown per image type.
\begin{table}[]
    \centering
    \begin{tabular}{c|c}
        \hline
        Type & Quantity \\
        \hline
         Bias & 922  \\
         Dark & 30  \\
         Flat & 1003  \\
         Object & 19210  \\
         Pointing & 537  \\
    \end{tabular}
    \caption{The breakdown per image type for the observations from 2017-01-11 to 2017-05-28.}
    \label{tab:imagetype}
\end{table}

The observations include two dither patterns, named D25 and D26, which were changed from D26 to D25 on 2017-03-20, where there is only a slight difference between the patterns. The dithering shifted the telescope's pointing slightly in random pixel directions during exposures to slightly smear the PSF, which mitigates pixel effects.
The observations includes three fields: G262 (centered at $\alpha=05^h 00^m, \delta=22^\circ30'$), G139 (centered at $\alpha=10^h 12^m$, $\delta=45^\circ00'$), and G111 (centered at $\alpha=16^h 40^m$, $\delta=52^\circ30'$). We manually checked each night's images to determine which frames were worthy of processing (i.e, no star smearing, no clouds, no weird beams of light, etc.). A breakdown of the number of object images per field and dither pattern is shown in Table \ref{tab:object_totals}.

\begin{table}[]
    \centering
    \begin{tabular}{c|c|c}
        \hline
        Field & D25 & D26 \\
        \hline
         G262 & 873 & 565  \\
         G139 & 6184 & 1257  \\
         G111 & 6477 & 405
    \end{tabular}
    \caption{The breakdown of total object images for each field and dither pattern for the observations.}
    \label{tab:object_totals}
\end{table}

\subsection{Results}
\label{s:results}
Using AutoWISP, following the procedure described in Section \ref{c:methodology}, for each image, we gather brightness measurements for each star with Gaia magnitude ${G}<{12}$, which astrometry projects within the image.

\begin{figure*}
    \centering
    \includegraphics[width=0.45\textwidth]{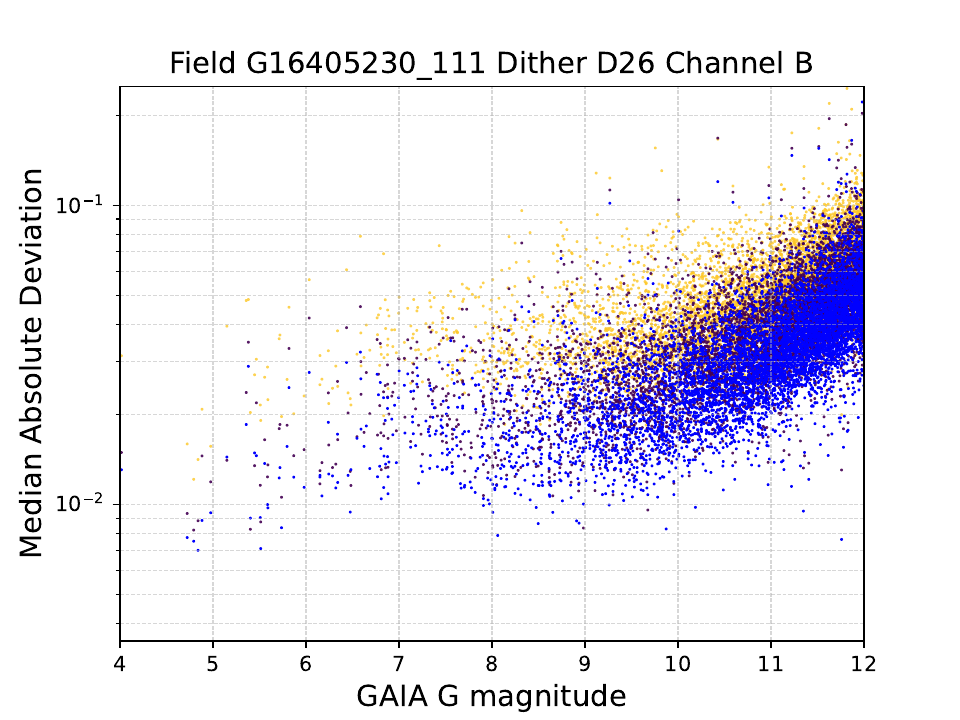}{(a)}
    \includegraphics[width=0.45\textwidth]{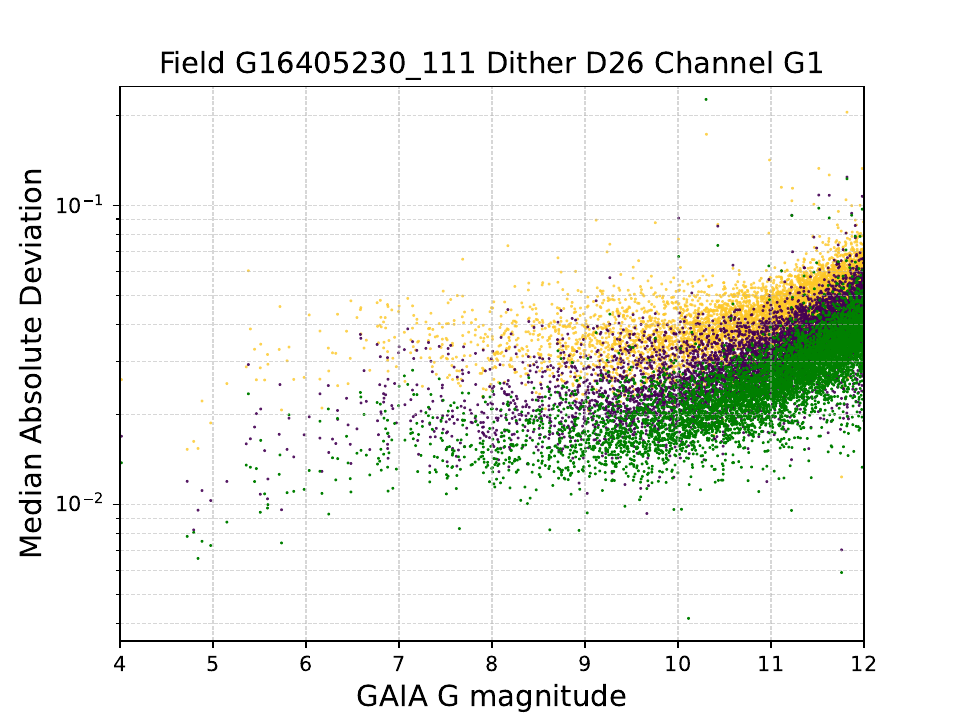}{(b)}
    \includegraphics[width=0.45\textwidth]{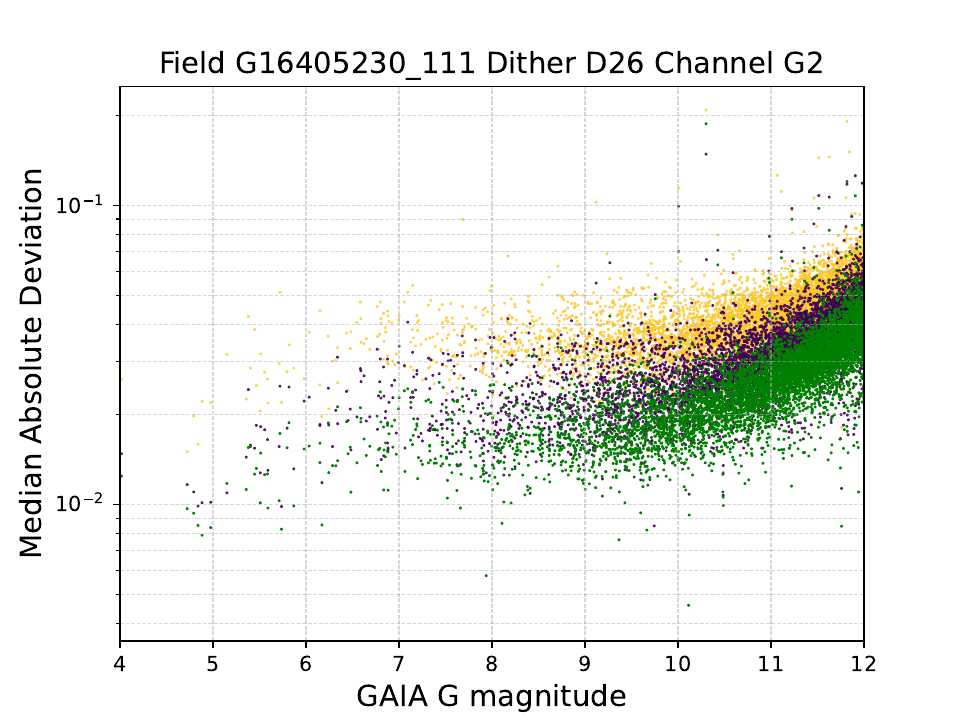}{(c)}
    \includegraphics[width=0.45\textwidth]{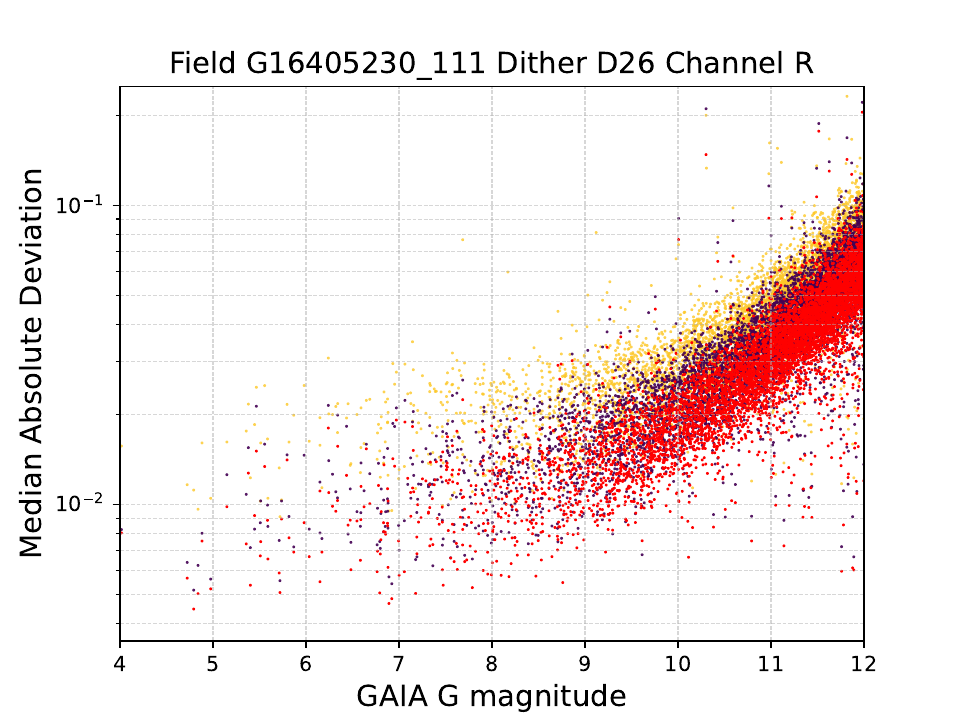}{(d)}
    \caption{The scatter (median absolute deviation from the median (MAD)) of the individual channel light curves vs. GAIA G magnitude before EPD (only magnitude-fitting) (indicated by yellow points), after EPD but before TFA (indicated by purple points), and after TFA (indicated by their corresponding channel color (B (blue)(a), G1 (green)(b), G2 (green)(c), R (red)(d)) points). These plots are for dither pattern D26, field G111. These show the overall improvements from our post-processing steps.}
    \label{fig:G111madD26}
\end{figure*}

\begin{figure*}
    \centering
    \includegraphics[width=0.45\textwidth]{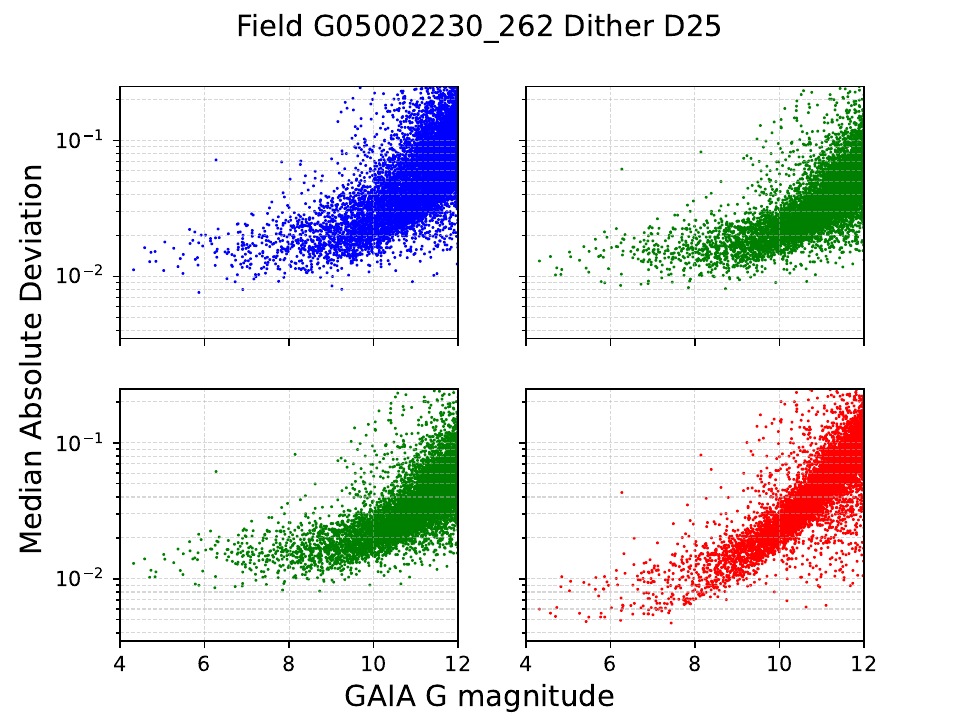}{(a)}
    \includegraphics[width=0.45\textwidth]{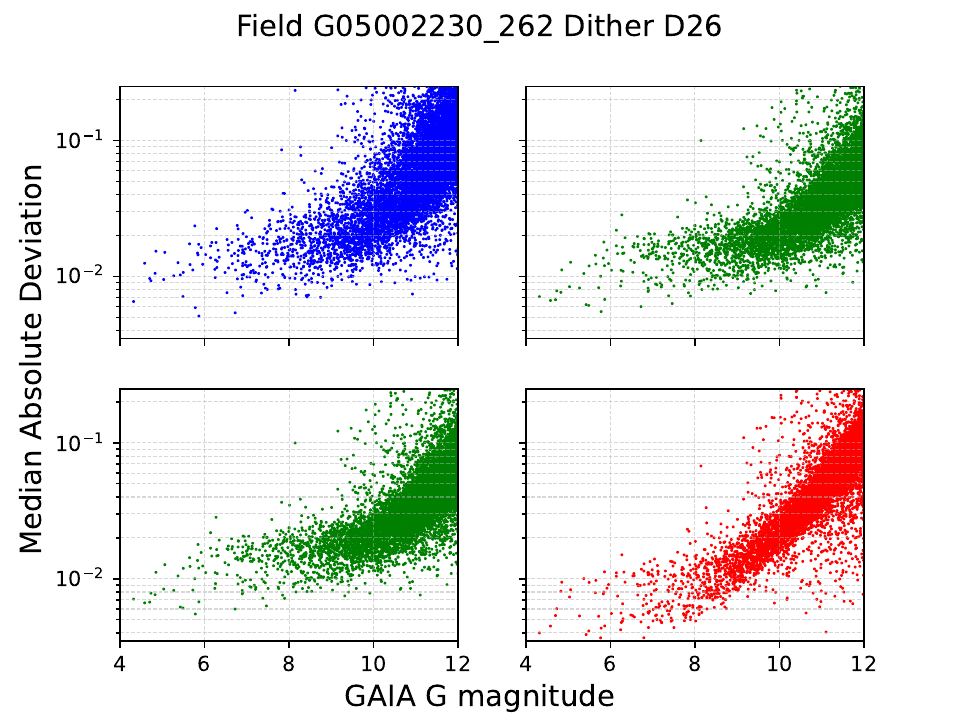}{(b)}
    \includegraphics[width=0.45\textwidth]{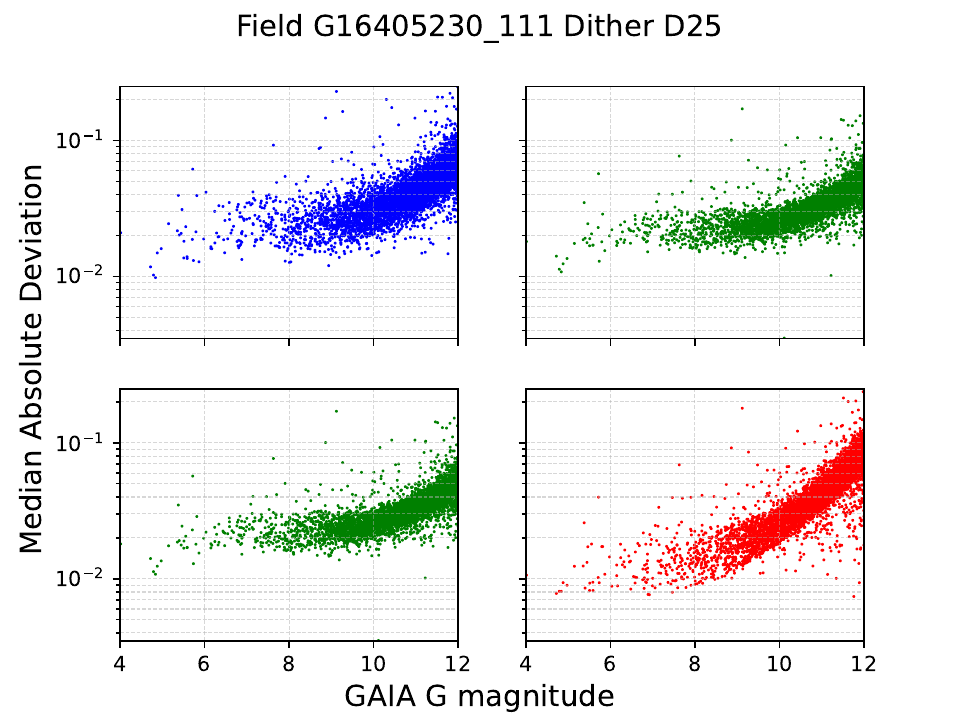}{(c)}
    \includegraphics[width=0.45\textwidth]{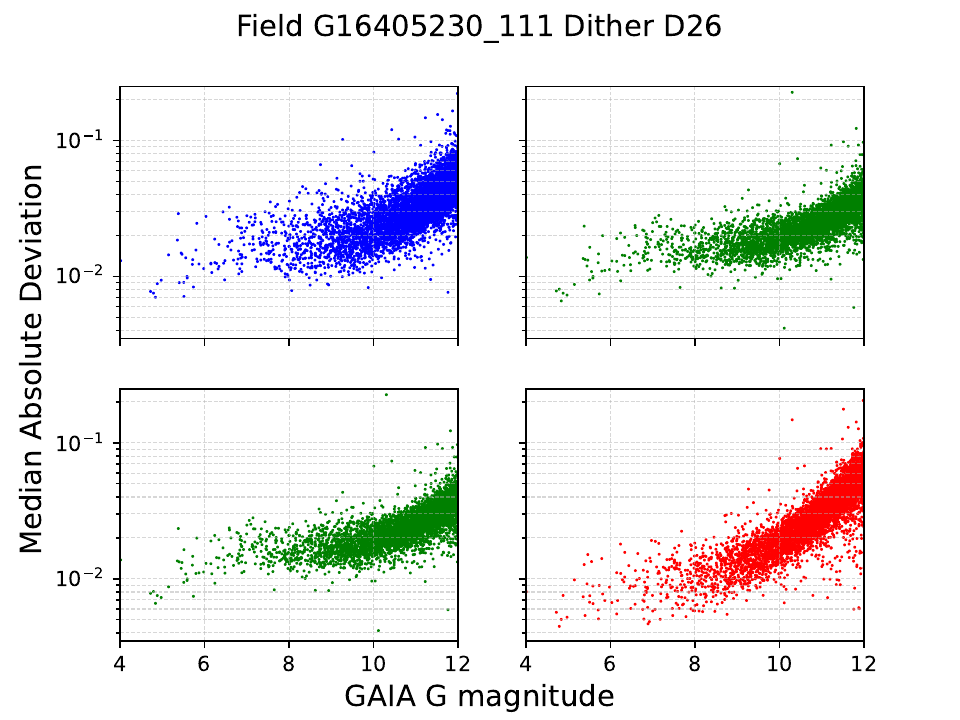}{(d)}
    \includegraphics[width=0.45\textwidth]{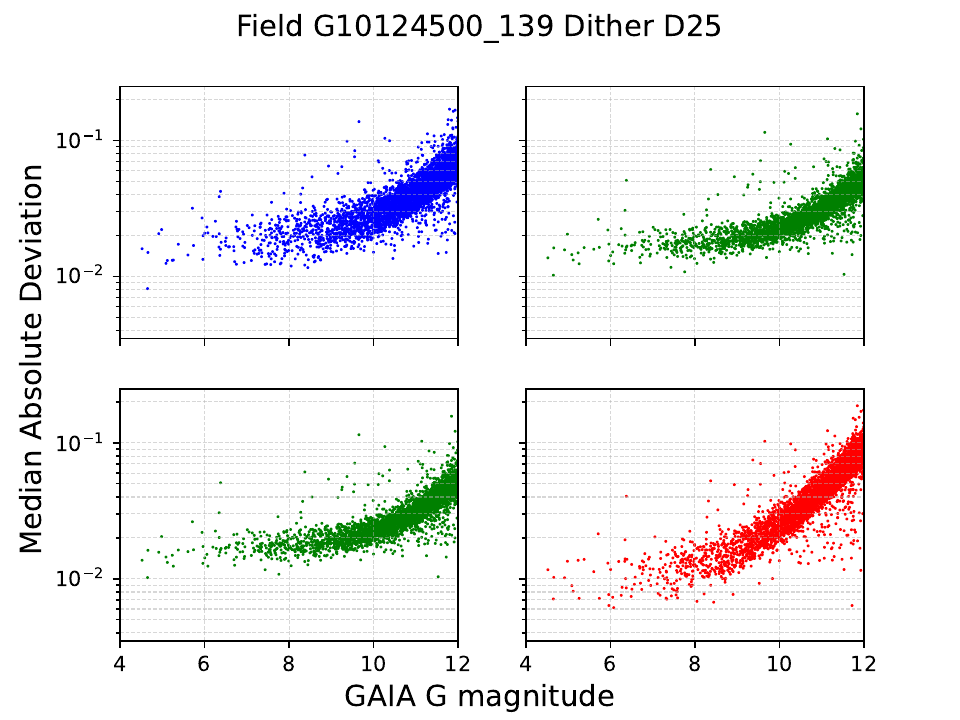}{(e)}
    \includegraphics[width=0.45\textwidth]{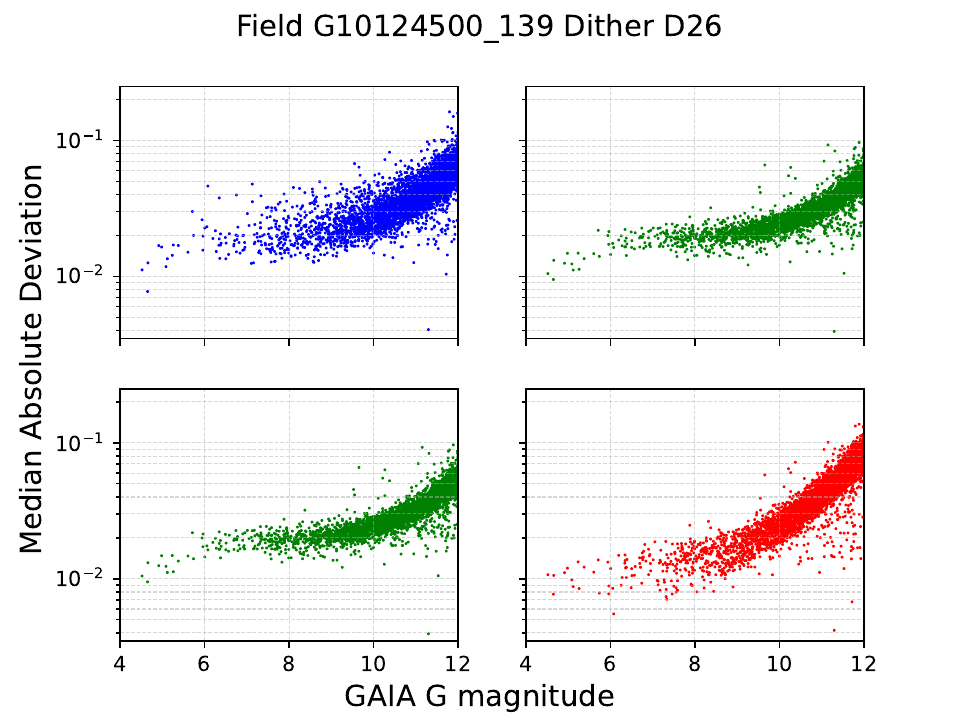}{(f)}
    \caption{The scatter (median absolute deviation from the median (MAD)) of the individual channel completely processed (i.e, after performing magnitude fitting, EPD, and TFA steps) light curves vs. GAIA G magnitude (indicated by their corresponding channel color (i.e., B (blue), G1 (green), G2 (green), R (red)). These plots are for dither patterns D25 (left plots (a),(c),(e)) and D26 (right plots (b),(d),(f)), for each field G262(a, b), G111 (c,d), and G139 (e,f)}.
    \label{fig:madD25D26}
\end{figure*}

\begin{figure*}
    \centering
    \includegraphics[width=0.45\textwidth]{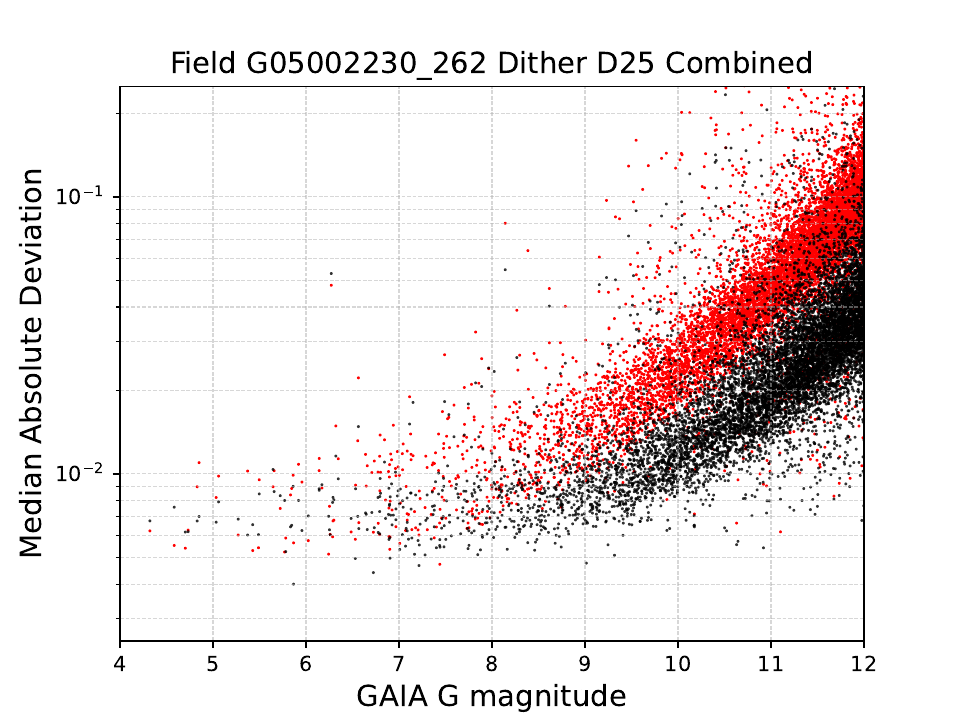}{(a)}
    \includegraphics[width=0.45\textwidth]{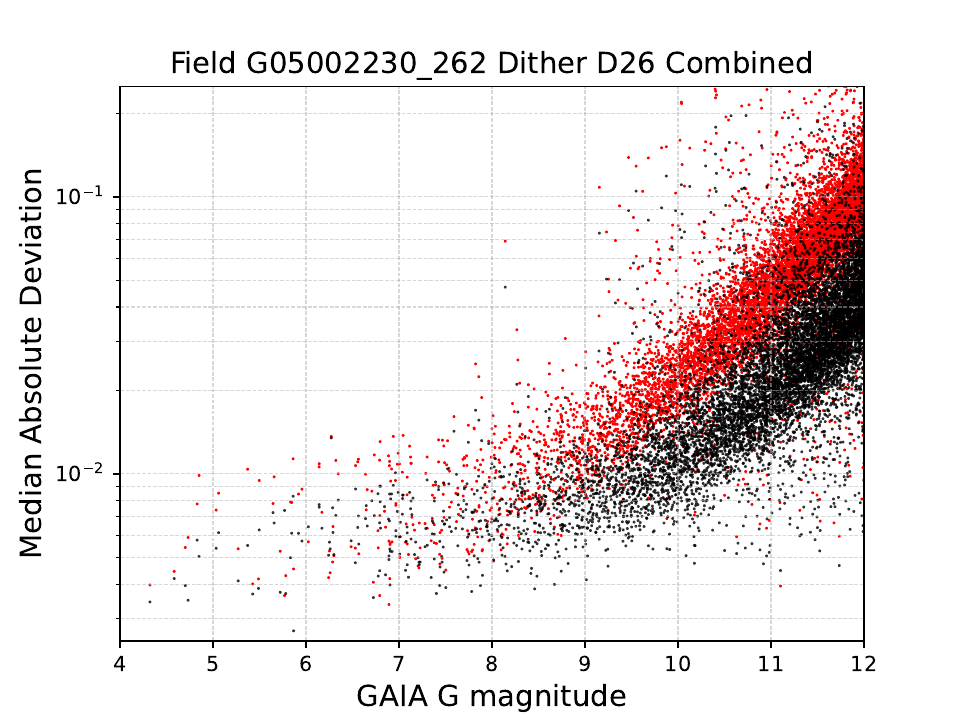}{(b)}
    \includegraphics[width=0.45\textwidth]{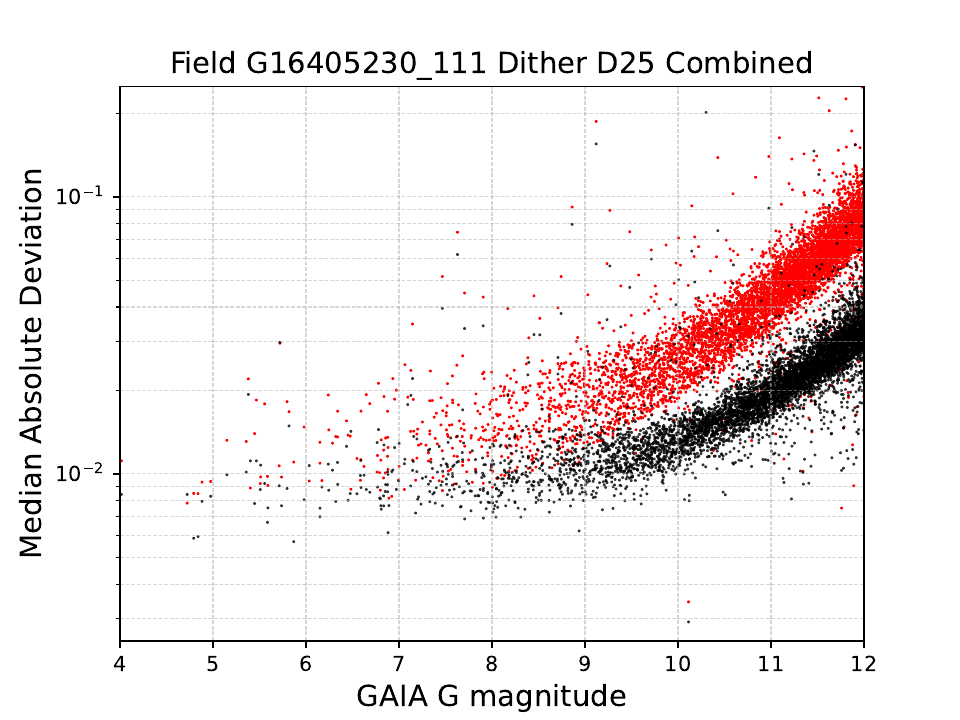}{(c)}
    \includegraphics[width=0.45\textwidth]{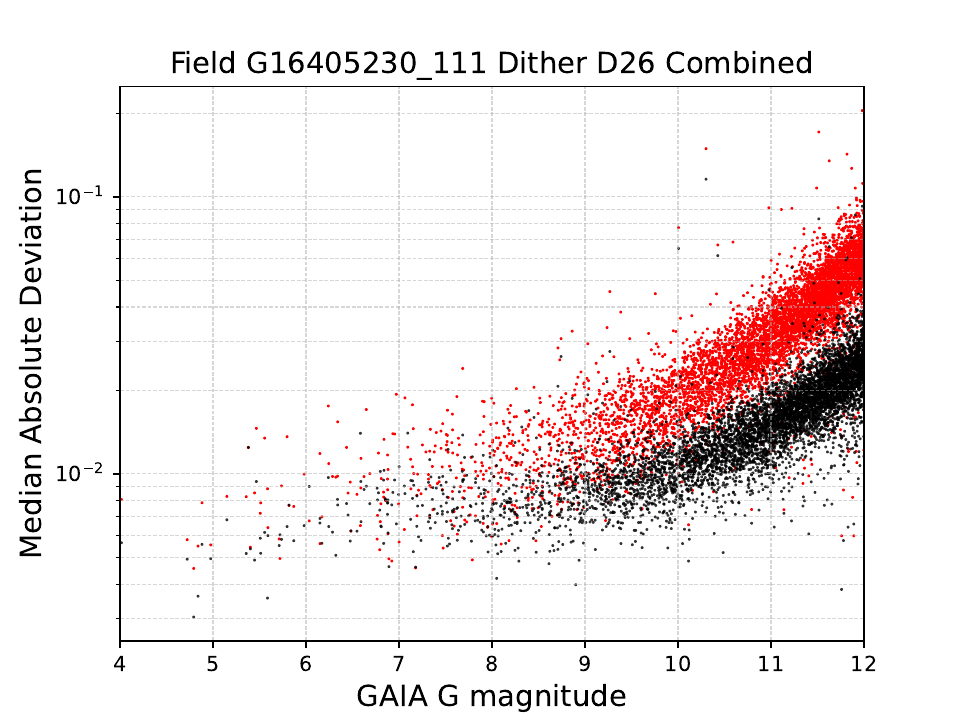}{(d)}
    \includegraphics[width=0.45\textwidth]{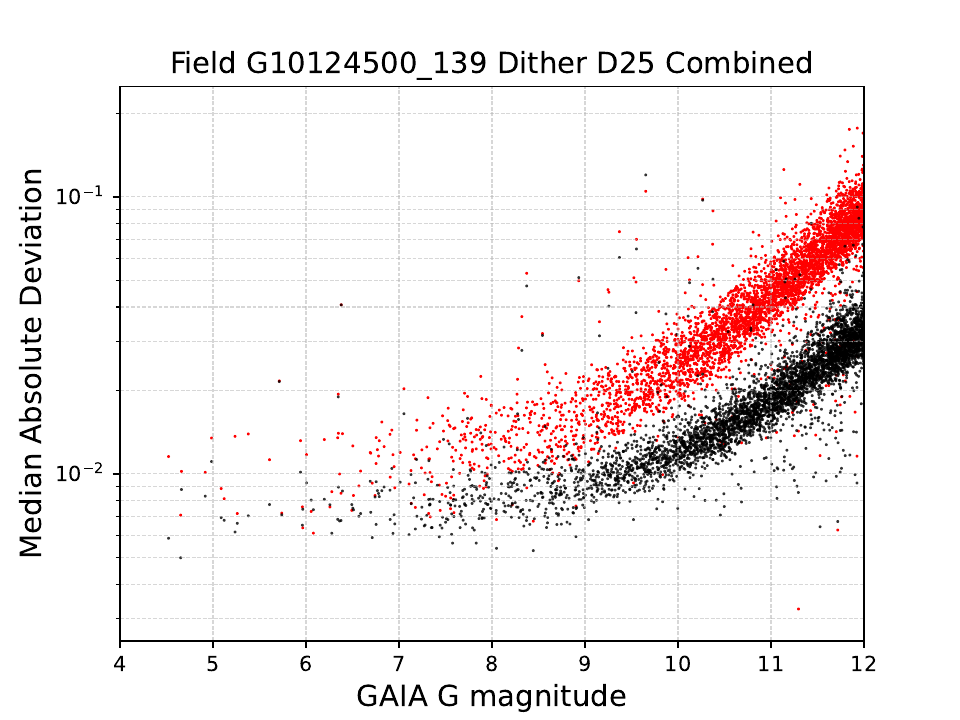}{(e)}
    \includegraphics[width=0.45\textwidth]{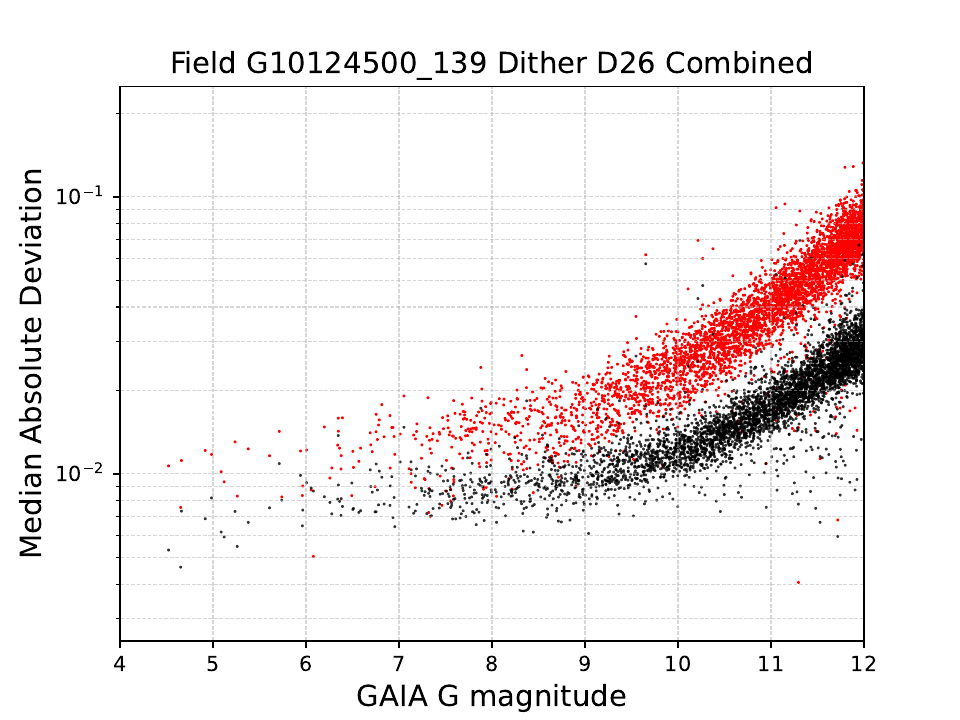}{(f)}
    \caption{The scatter (median absolute deviation from the median (MAD)) of the combined channel completely processed (i.e, after performing magnitude fitting, EPD, and TFA steps) light curves vs. GAIA G magnitude (black points) for fields G262 (a,b), G111 (c,d), and G139 (e,f) for dither patterns D25 (left) and D26 (right). For comparison, the red channel scatter is also shown (red points)}
    \label{fig:combinedmadD25D26}
\end{figure*}

To demonstrate the photometric precision, we plot the median absolute deviation (median of the absolute value of the difference from the median) (MAD) versus the GAIA catalogue G magnitude for each star in the frame in Figures \ref{fig:madD25D26}, \ref{fig:G111madD26}, and \ref{fig:combinedmadD25D26}. Note that MAD is smaller than the standard deviation by a factor of 1.48 for Gaussian distributions.

%

\begin{figure}
    \centering
    \includegraphics[width=0.45\textwidth]{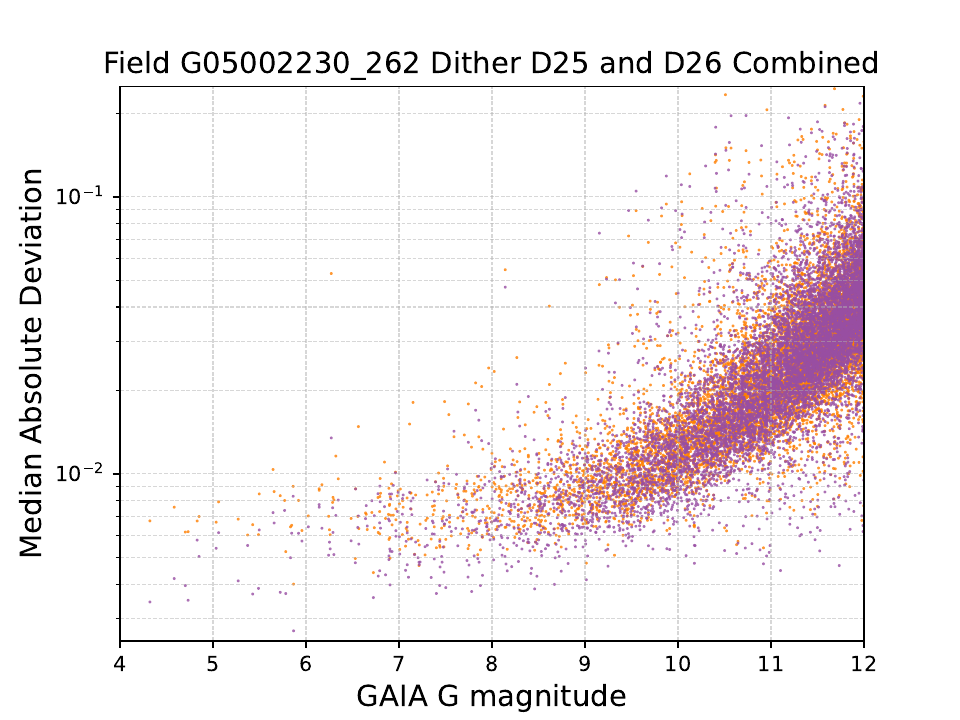}{(a)}
    \includegraphics[width=0.45\textwidth]{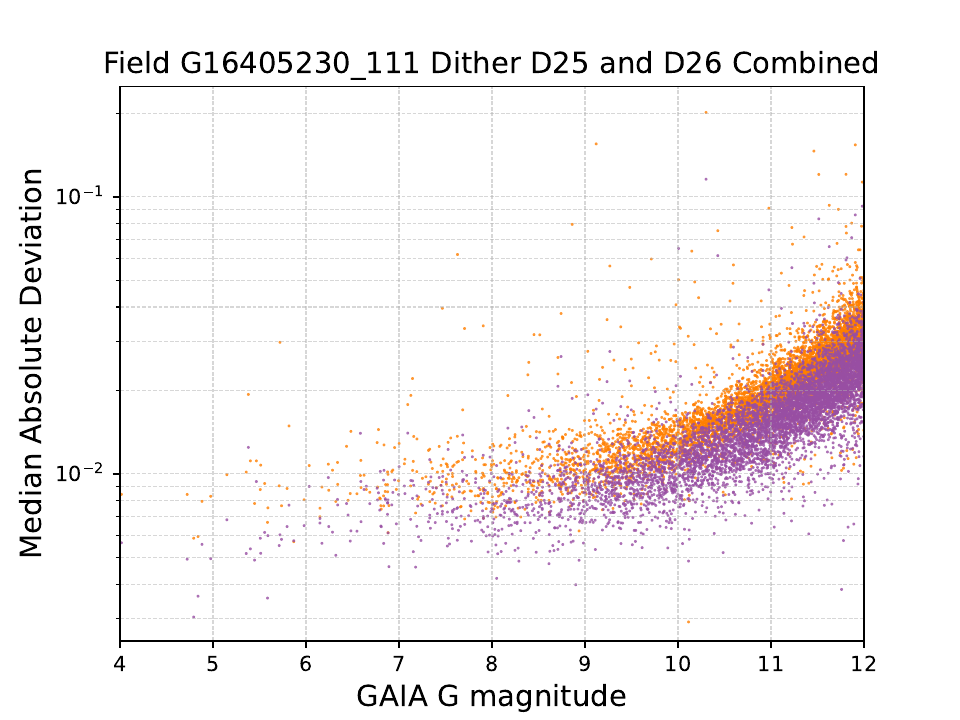}{(b)}
    \includegraphics[width=0.45\textwidth]{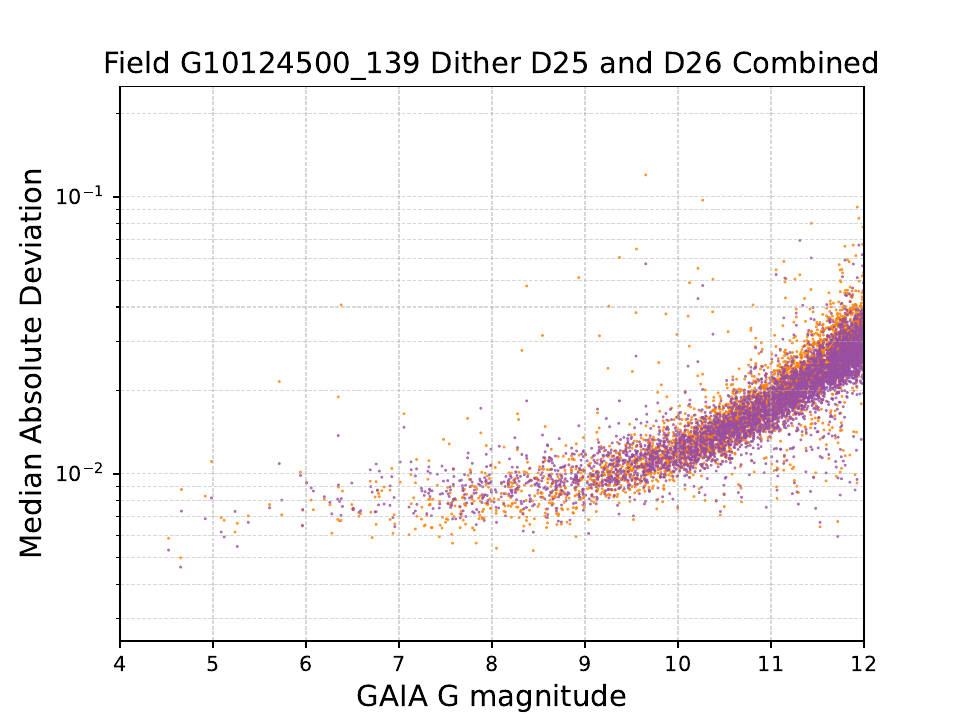}{(c)}
    \caption{Similar to Figure \ref{fig:combinedmadD25D26}, comparing the scatter of the two dither patterns, D25 (orange points) and D26 (purple points).}
    \label{fig:comparedmadD25D26}
\end{figure}

This measure of performance is relatively insensitive to outliers. Our measurements' precision is typical for photometry, with faint stars having a high MAD and very bright stars having a low MAD, where faint stars have high sky and photon noise that are predominant over their measurements. Typically, very bright stars have oversaturated pixels, making it hard to measure their flux precisely, but in our case, we do not seem to hit our saturation limit. Figure \ref{fig:madD25D26} shows each field, dither pattern, and channel precision after being completely processed by our pipeline (i.e, after performing magnitude fitting, EPD, and TFA steps). Figure \ref{fig:G111madD26} demonstrates the increased performance from the magnitude fitting, EPD, and TFA steps for field G111, dither pattern D26. Figure \ref{fig:combinedmadD25D26} shows the combined color channels' precision after being completely processed by our pipeline as described in Sections \ref{s:calibration}-\ref{s:post_processing} (i.e, after performing magnitude fitting, EPD, and TFA steps). Combining the channel scatters was done by incorporating a weighted average of each channel, with weights chosen to minimize the scatter (MAD). Comparing the two dither patterns used, the D26 dither pattern performed slightly better photometrically, achieving a slightly lower precision per field, as seen in Figure \ref{fig:comparedmadD25D26}.


We determined our scintillation noise limit using the modified Young approximation \citep{Young_67, Osborn_2015}, finding it to be 4 mmag $\sigma\approx$ 2.7 mmag MAD, at airmass 1.09. As a disclaimer, this scintillation noise limit is only an approximation for several reasons. First, the empirical correction coefficient $C_\gamma$ from the modified Young approximation was assumed to be 1.5 rather than being specifically determined for the observing site FLWO. Additionally, we are uncertain how the different color filters (R, G, B) may have affected this empirical correction. Finally, we had to extrapolate the modified Young approximation, as it is primarily defined for larger telescopes. Using the combined precisions from Figure \ref{fig:combinedmadD25D26}, we do not hit this limit. Furthermore, we determined our saturation limit to be Gaia G${\approx 5.5}$ mag for each color channel. This was done by taking the maximum pixel value for Gaia ${G}<{7}$ mag stars in all the raw frames and comparing when a large percentage of those corresponding stars went over our saturation threshold of 14000 ADU.

\subsection{Example light curves:}
\label{s:example_lcs}
All of the data was searched for transiting exoplanets and eclipsing binaries, cross matching GAIA identifiers to the exoplanet archive, Kepler \citep{Brown_et_al_11}, K2 \citep{Kirk_et_al_16}, Pr{\v{s}}a \citep{Prsa_et_al_11}, and GAIA eclipsing binaries \citep{Mowlavi_et_al_23} catalogues.

Of the transiting exoplanet light curves found in each field, the photometric signal was either too weak to detect their transits, or the transit did not occur during the set of observations.

We plot the change in magnitude versus phase periods (one cycle of the period folded) for the corresponding GAIA identifiers (titles of the plots) of several eclipsing binaries, as shown in Figures \ref{fig:GDR3_1415133175038821888}, \ref{fig:GDR3_1425033869225305728}, \ref{fig:GDR3_1431129011933593088}, \ref{fig:GDR3_829685534384441344}, and \ref{fig:GDR3_829393545336090368}, with red, green, and blue for each color light curve, or teal if the colors were combined. The corresponding TESS light curves (black points), obtained from the TESS Data Pipeline (SPOC) via the MAST astroquery API \citep{Ginsburg_et_al_19} for each of their corresponding observation sectors, are overlaid on our data points. The simultaneous multi-color photometry provided by DSLR observations offers information on the wavelength dependence of any variability signal observed under identical conditions. This adds valuable insight to aid in the interpretation of the results.

Additionally, individual data points were binned (represented by circle points in the corresponding colors) to highlight the trend in the signal better. The light curves shown are after being completely processed (i.e, after performing magnitude fitting, EPD, and TFA steps). The periods used for each system was either found from the TESS input catalogue or by adjusting the phase period until our light curves data points and TESS light curves matched. All the TESS data used in this paper can be found in MAST: \dataset[https://doi.org/10.17909/t9-st5g-3177]{https://doi.org/10.17909/t9-st5g-3177} and \dataset[https://doi.org/10.17909/t9-nmc8-f686]{https://doi.org/10.17909/t9-nmc8-f686}.

Of the eclipsing binaries found, we include some that we found interesting, such as the following:

\subsubsection{GQ Dra}
\label{ss:GQ_Dra}
\begin{figure*}
    \centering
    \includegraphics[width=0.45\textwidth]{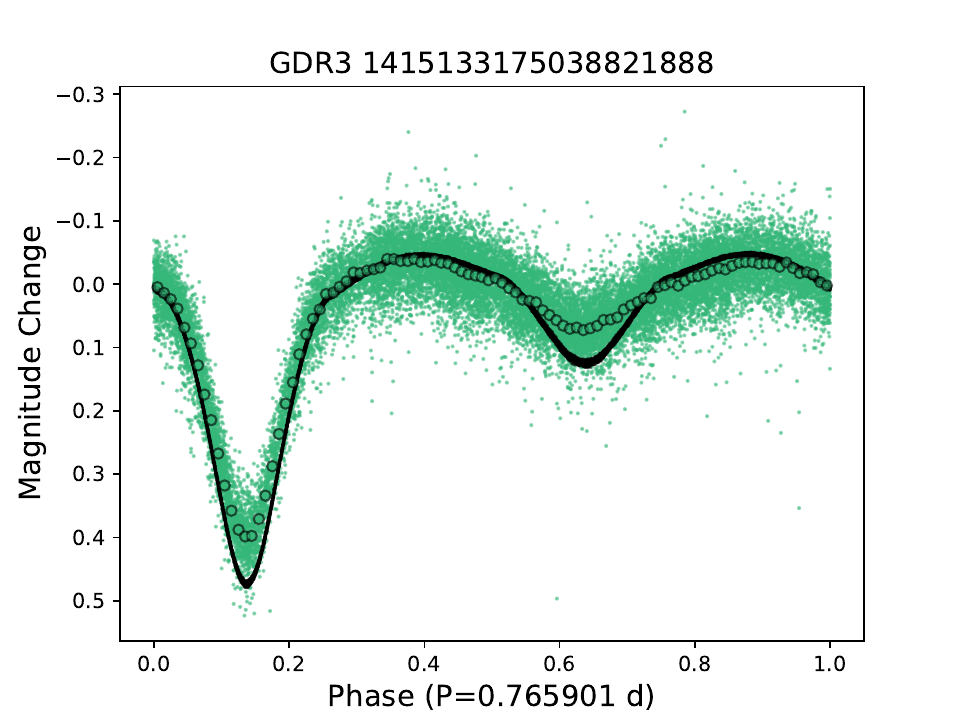}{(a)}
    \includegraphics[width=0.45\textwidth]{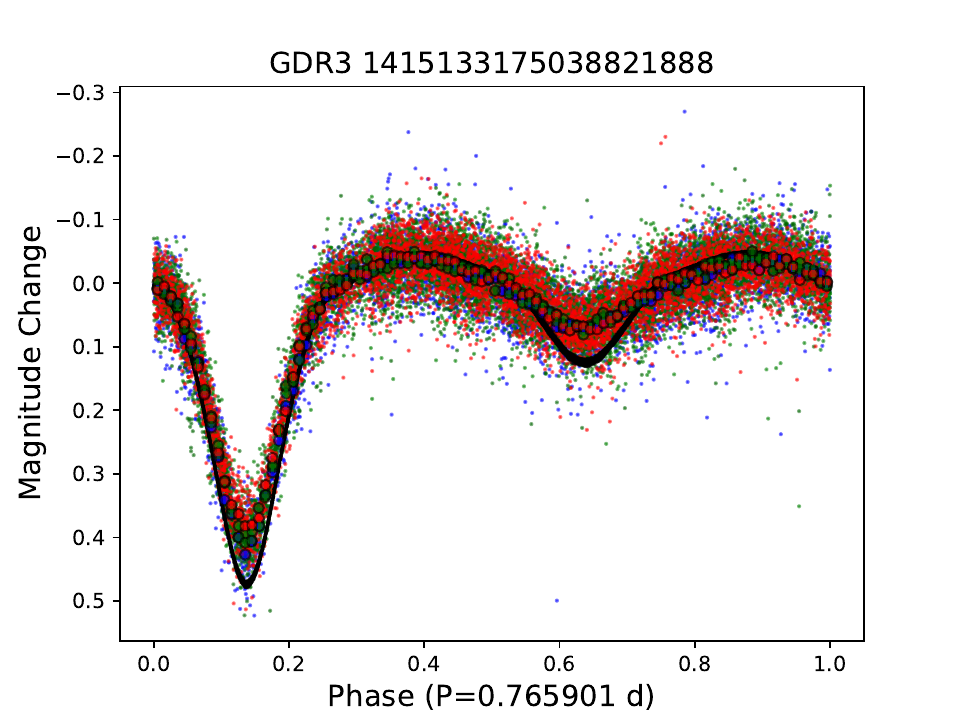}{(b)}
    \caption{The change in magnitude versus phase plot of GQ Dra (GDR3 1415133175038821888). We determined its period to be 0.765901 days from combining TESS and our data. (a.) Our combined light curve points (teal) are binned (circle teal points) and folded by the period. The corresponding TESS light curves (sectors 74, 78, 80, and 84) (black points) are included. (b.) The binary system showed no color dependence (no obvious variance in red, blue, or green data points).}
    \label{fig:GDR3_1415133175038821888}
\end{figure*}

Figure \ref{fig:GDR3_1415133175038821888} shows the phase plot of GQ Dra (GDR3 1415133175038821888, Gaia G magnitude 9.1891). GQ Dra is an eclipsing binary system with at least one pulsating component. It is classified as a $\delta$ Scuti and $\beta$ Lyrae type variable. It is a semi-detached binary system with a hot, pulsating primary and a cooler secondary component \citep{Ulsas_et_al_20}, the latter (fainter) filling its Roche lobe. We determined the period of this binary system by combining our observations with the TESS light curve and found that it has a period of 0.765901 days. GQ Dra's variability was first discovered by the Hipparcos satellite \citep{ESA_97} and has been extensively characterized by many researchers \citep{Atay_et_al_00, Qian_et_al_15, Ulsas_et_al_20, Liakos_et_al_16}.

From Figure \ref{fig:GDR3_1415133175038821888}, we observe that the eclipses in our light curves are much shallower compared to those from TESS (sectors 74, 78, 80, and 84). Additionally, our period is slightly different from the value found in the literature (0.7659029(2) days \citep{Ulsas_et_al_20}). These variations in eclipse depth may be because our observations (from 2017) are nearly 7 years apart from the TESS observations (sector 74 in January 2024 and sector 84 between October 2024). The underlying mechanism might operate on a timescale of years, meaning we may not have captured it due to gaps in our observations. Furthermore, there is no indication of color dependence, so the difference in eclipse depth is unlikely to be explained by the different passbands of the Sony-$\alpha$ sensor versus TESS (as described in Section \ref{ss:TX_UMa}). Comparing the light curve for this system after the magnitude fitting, EPD, and TFA steps, no change in eclipse depth was observed, so neither EPD nor TFA distorted the target signal. Checking the Gaia catalogue for nearby stars to GQ Dra, we found one star (GDR3 1415133179334896128) with a Gaia G magnitude 10.952358 within 5 arc seconds of GQ Dra. This star is close and bright enough to blend with GQ Dra. Ultimately, TESS corrects for the blending of close stars, which may also be the reason why our eclipse depths are shallower.

\vfill

\subsubsection{GG Dra}
\label{ss:GG_Dra}
\begin{figure*}
    \centering
    \includegraphics[width=0.45\textwidth]{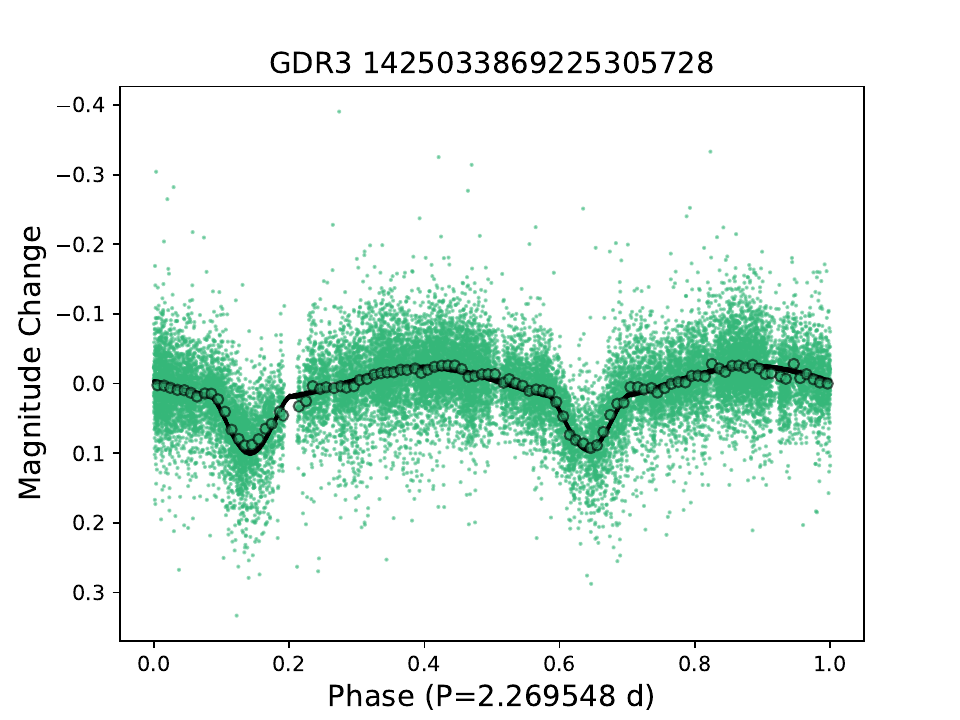}{(a)}
    \includegraphics[width=0.45\textwidth]{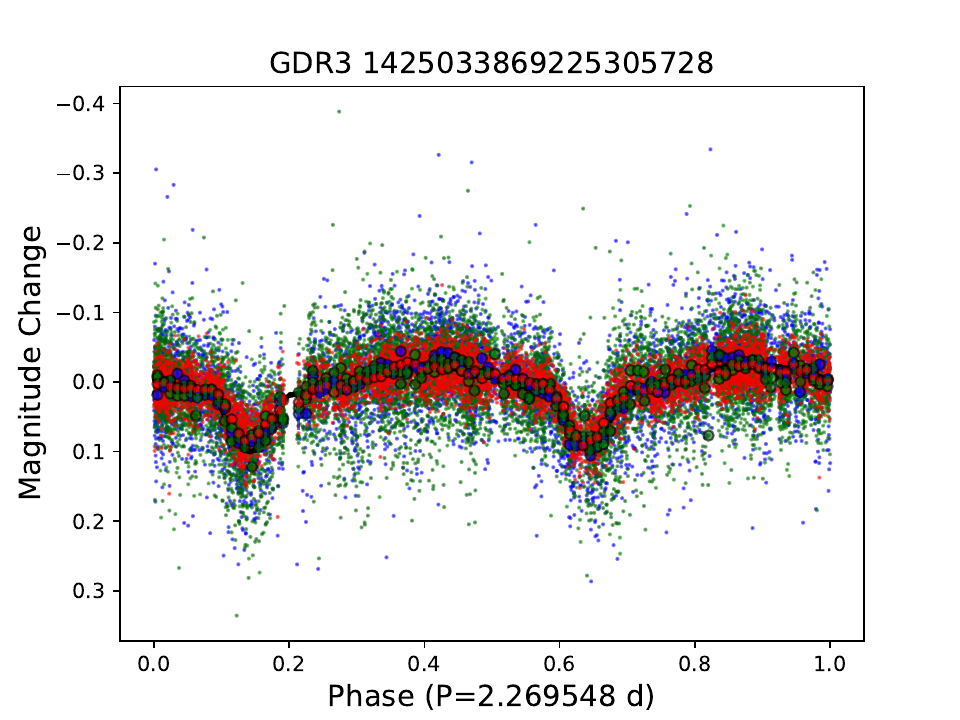}{(b)}
    \caption{The change in magnitude versus phase plot of GG Dra (GDR3 1425033869225305728, Gaia G magnitude 8.6141). We determined its period to be 2.2695484 days from combining TESS and our data. (a.) Our light curve points (teal) are binned (circle teal points) and folded by the period. The corresponding TESS light curves (sectors 16, 23, 25, 50, 51, 52, 77, 78, 79, and 83) (black points) are included. (b.) The binary system showed no color dependence (no obvious variance in red, blue, or green data points).}
    \label{fig:GDR3_1425033869225305728}
\end{figure*}

Figure \ref{fig:GDR3_1425033869225305728} shows the phase plot of GG Dra (GDR3 1425033869225305728, Gaia G magnitude 8.6141). GG Dra is a $\beta$ Lyrae-type eclipsing binary system with a period of 2.2695484 days. Like GQ Dra, GG Dra's variability was discovered by the Hipparcos satellite \citep{ESA_97}. This binary system, in particular, has not been extensively observed or characterized.

As seen in Figure \ref{fig:GDR3_1425033869225305728}, the depths of the primary and secondary are almost equal and/or differ very negligibly, and the period we determined (2.2695484 days) was found by adjusting the TESS light curves (sectors 16, 23, 25, 50, 51, 52, 77, 78, 79, and 83) until they overlapped equally. Our light curves match the TESS light curves reasonably well, showing no color dependence.

\subsubsection{V0551 Dra}
\label{ss:V0551_Dra}
\begin{figure*}
    \centering
    \includegraphics[width=0.45\textwidth]{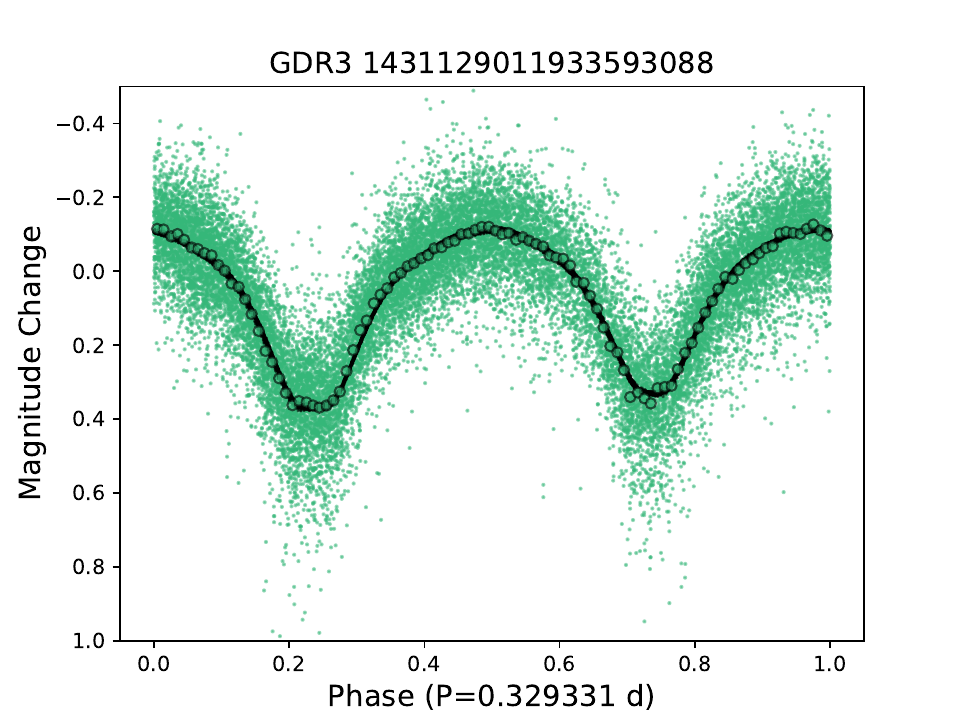}{(a)}
    \includegraphics[width=0.45\textwidth]{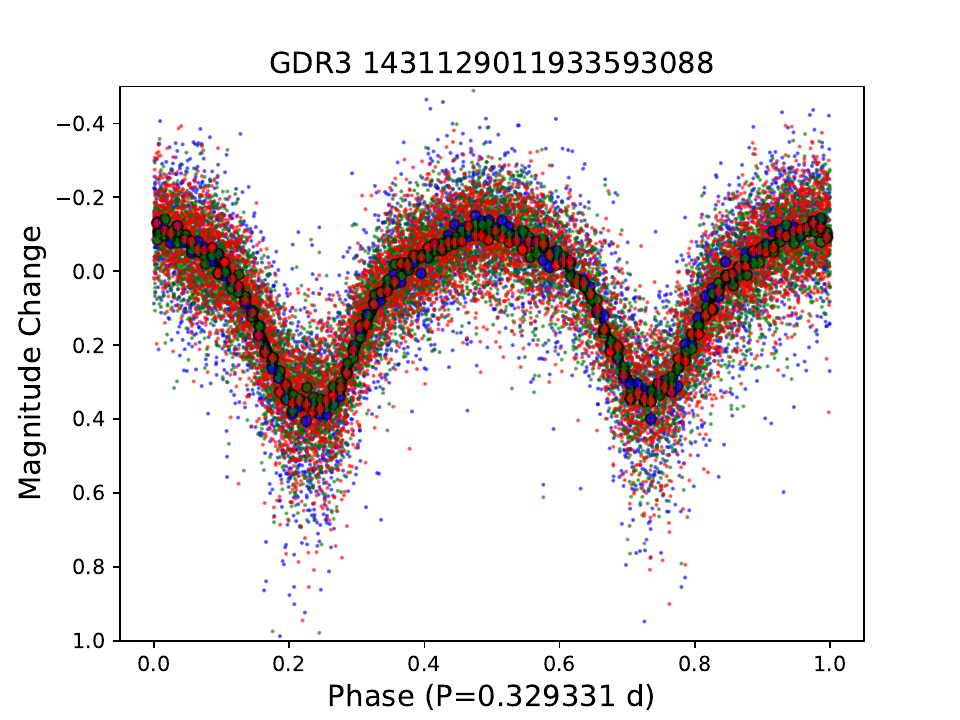}{(b)}
    \caption{The change in magnitude versus phase plot of V0551 Dra (GDR3 1431129011933593088, Gaia G magnitude 11.5235). We determined its period to be 0.329331 days from combining TESS and our data. (a.) Our light curve points (teal) are binned (circle teal points) and folded by the period. The corresponding TESS light curves (sectors 15, 16, 19, 22, 23, 24, 25, 49, 50, 51, 52, 56, 59, 76, 77, 78, 79, 83, and 86; black points) are included. (b.) The binary system showed no color dependence (no obvious variance in red, blue, or green data points).}
    \label{fig:GDR3_1431129011933593088}
\end{figure*}

Figure \ref{fig:GDR3_1431129011933593088} shows the phase plot of V0551 Dra (GDR3 1431129011933593088, Gaia G magnitude 11.5235). V0551 Dra is a W Ursae Majoris-type contact eclipsing binary. These are two strongly interacting ellipsoidal component stars with very short periods, nearly in contact with each other, filling their critical Roche lobes and sharing a common envelope \citep{Qian_et_al_20}.

As seen in Figure \ref{fig:GDR3_1431129011933593088}, the depths of the primary and secondary are almost equal and/or differ very negligibly, and the period we determined (0.329331 days), by combining our light curves with TESS, closely matches what is found in literature (0.3293313 days \citep{Kjurkchieva_et_al_18, Christopoulou_et_al_22}). TESS light curves (sectors 15, 16, 19, 22, 23, 24, 25, 49, 50, 51, 52, 56, 59, 76, 77, 78, 79, 83, and 86) are overlaid, where our light curves closely match TESS.

\subsubsection{TX UMa}
\label{ss:TX_UMa}
\begin{figure*}
    \centering
    \includegraphics[width=0.45\textwidth]{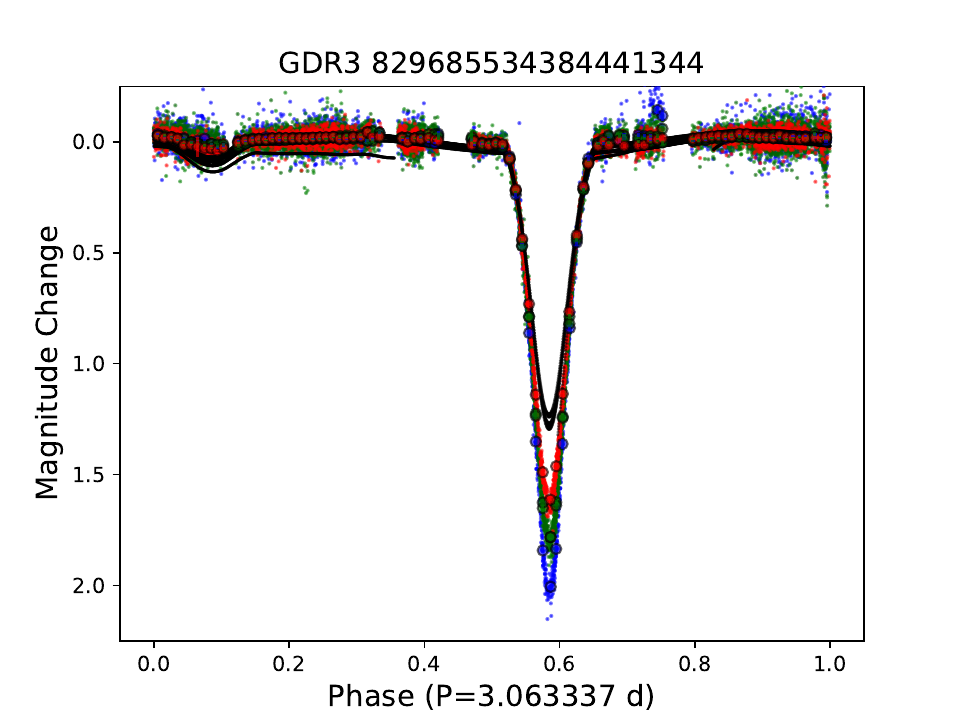}{(a)}
    \includegraphics[width=0.45\textwidth]{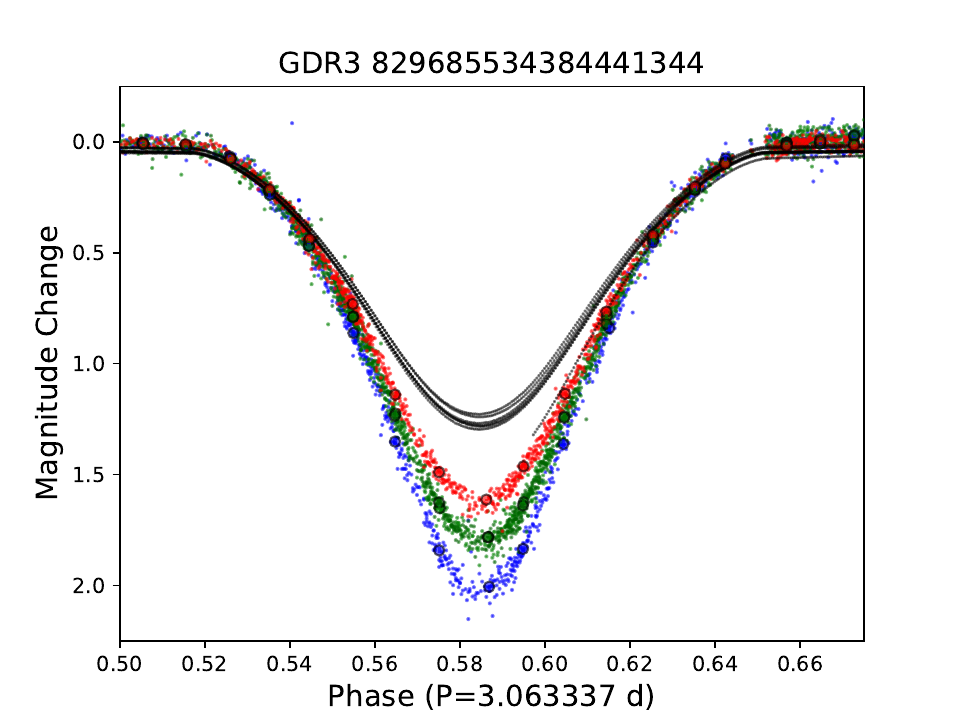}{(b)}
    \caption{The change in magnitude versus phase plot of TX Ursae Majoris (TX UMa) (GDR3 829685534384441344, Gaia G magnitude 6.9229). We determined its period to be 3.0633368 days from combining TESS and our data. (a.) Our light curve points (corresponding to each color channel (i.e., red, green, and blue)) are binned (each corresponding color channel (i.e., red, green, and blue) circle points) and folded by the period. The corresponding TESS light curves (sectors 21 and 48) (black points) are included. (b.) The zoomed-in plot shows the clear changes in the light curves per color channel and changes in each TESS light curve.}
    \label{fig:GDR3_829685534384441344}
\end{figure*}

Figure \ref{fig:GDR3_829685534384441344} shows the phase plot of TX Ursae Majoris (TX UMa) (GDR3 829685534384441344, Gaia G magnitude 6.9229). TX UMa is a semi-detached $\beta$ Persei-type (Algol) binary system that exhibits period variations. It was first discovered in 1925 as a single-line spectroscopic binary by J. A. Pearce and later identified as an eclipsing system by H. R{\"u}gemer and H. Schneller in 1931 \citep{Komzik_et_al_92, Glazunova_et_al_11}. TX UMa is a well-studied main-sequence binary system \citep[e.g.,][]{Hiltner_45, Plavec_60, Kreiner_et_al_80, Grygar_et_al_91, Qian_01, Glazunova_et_al_11}. It consists of a B8 V spectral type main-sequence primary star and an G0 III–IV spectral type evolved giant secondary \citep{Cester_et_al_77, Alard_Lupton_98, Glazunova_et_al_11}. The primary star has a temperature of 12,900 K, making it a very hot blue star. It exhibits a changing eclipse period, which was eventually explained by apsidal precession \citep{Plavec_60}. The secondary component is a cool, evolved giant with a temperature of 5,500 K. Having exhausted its core hydrogen, it has evolved off the main sequence. Its Roche lobe is filled, contributing mass to the primary \citep{Glazunova_et_al_11}.

Figure \ref{fig:GDR3_829685534384441344} shows each color channel plotted against the phase, with TESS light curves (sectors 21 and 48) overlaid. There is a clear indication of color dependency, with the primary eclipse getting shallower with increasing wavelength. The secondary eclipse is detected with too low signal-to-noise to determine color dependency. Additionally, we observe that each TESS light curve changes in each sector. This could be due to a remaining long-term trend that causes variations in different transits, although the cause remains unclear. Our derived period of 3.06333(2) days agrees with the literature value of 3.0633292 days \citep{Malkov_20}. The different eclipse depths per color channel are attributed to the varying bandpasses used by the Sony-$\alpha$7R II DSLR camera and TESS, as shown in Figure \ref{fig:bandpass}.

\subsubsection{GW UMa}
\label{ss:GW_UMa}
\begin{figure}
    \centering
    \includegraphics[width=0.5\textwidth]{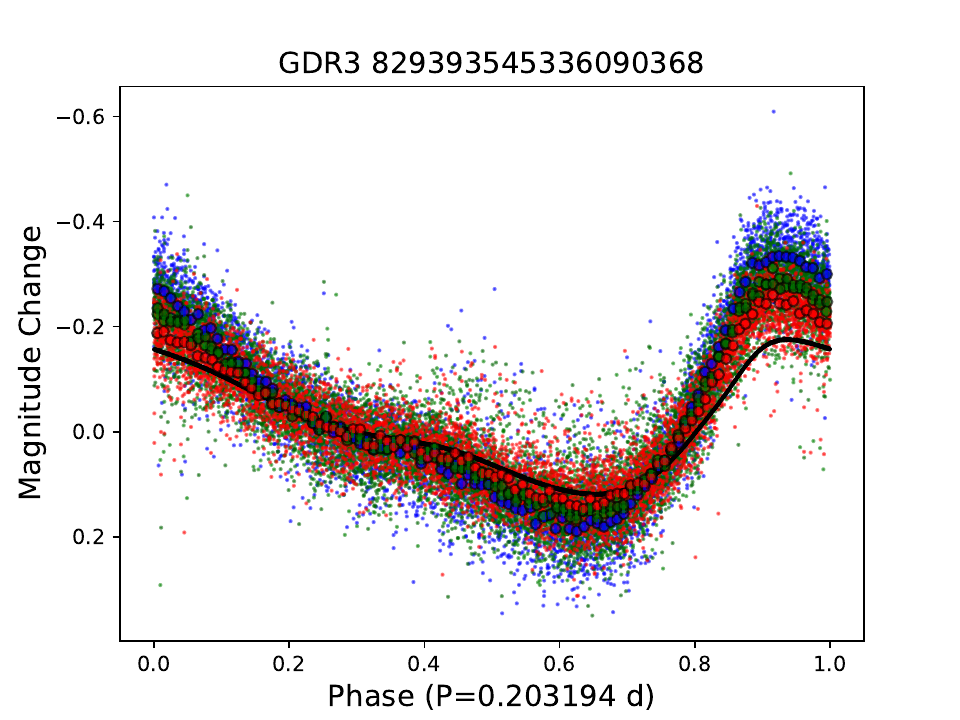}
    \caption{The change in magnitude versus phase plot of GW Ursae Majoris (GW UMa) (GDR3 829393545336090368, Gaia G magnitude 9.6529). We determined its period to be 0.203194 days from combining TESS and our data. Our light curve points (corresponding to each color channel (i.e., red, green, and blue)) are binned (each corresponding color channel (i.e., red, green, and blue) circle points) and folded by the period. The corresponding TESS light curves (sectors 21 and 48) (black points) are included.}
    \label{fig:GDR3_829393545336090368}
\end{figure}

Figure \ref{fig:GDR3_829393545336090368} shows the phase plot of GW Ursae Majoris (GW UMa) (GDR3 829393545336090368, Gaia G magnitude 9.6529). GW UMa is a stable, monoperiodic, high-amplitude $\delta$ Scuti star \citep{Derekas_et_al_03, Hintz_et_al_05} (spectral type A8 IV) discovered by the Hipparcos satellite, with a period of 0.2032 days \citep{ESA_97} (revised period of 0.20319389 days) \citep{Hintz_et_al_05}.

As seen in Figure \ref{fig:GDR3_829393545336090368}, each color channel is plotted against the phase, with TESS light curves (sectors 21 and 48) overlaid. Our period (0.203194 days) matches precisely with the value found in the literature (0.20319389 days \citep{Hintz_et_al_05}), and there is clear evidence of color (wavelength) dependence, as discussed in Section \ref{ss:TX_UMa}, which is typical of many $\delta$ Scuti variables.

\vfill
\section{Discussion}
\label{c:discussion}
This article describes an automated pipeline, AutoWISP, to perform photometry on astronomical images. AutoWISP consists of six major steps: image calibration, plate-scale solving, photometry, magnitude fitting, creating light curves, and light curve post-processing. We demonstrate the pipeline's capability using a dataset from a Sony-$\alpha$7R II (ILCE-7RM2) DSLR Camera attached to HAT10, a unit of HATNet at FLWO. We fully reduced this dataset using AutoWISP, from raw images to light curves, showcasing light curves of five example eclipsing binary systems. Through these efforts to create an automated photometry process, we aim to dramatically enhance the scientific return of citizen scientist observations and significantly expand the pool of citizen scientists by enabling high-precision photometry with ubiquitous, low-cost color cameras. Furthermore, color detectors provide valuable simultaneous color information for stars and planets. For example, in the case of eclipsing binaries, color information can help estimate the temperature difference between the stars, provide additional constraints on their sizes and limb darkening effects, and reveal insights into their atmospheric properties. Lastly, our goal is that by delivering AutoWISP to the citizen science community, we will simultaneously increase the number of observers capable of photometric follow-up and expand the range of targets they can pursue. This will ultimately allow for more extended time baselines than those possible with individual surveys alone (such as TESS), enabling professionals and citizen scientists to observe objects more frequently and at later times without initiating entirely new missions.

\subsection{Project PANOPTES}
\label{s:panoptes}
We are currently collaborating with Project PANOPTES (Panoptic Astronomical Networked Observatories for a Public Transiting Exoplanets Survey), a global scientific network of cameras searching for habitable planets and is described as 'a citizen science project that aims to make it easy for anyone to build a low-cost robotic telescope that can be used to detect transiting exoplanets' \citep{Guyon_et_al_14}. Using this pipeline, as described in Section \ref{c:methodology}, and having already achieved sub-percent photometry per channel for Project PANOPTES using our AutoWISP pipeline, we will produce fully processed, high-precision light curves for their entire network of telescopes.

\subsection{Challenges}
\label{s:challenges}
For the Sony-$\alpha$ dataset, we attempted to model the PSF/PRF for the entire dataset to achieve the best possible photometry. We tested different modeling procedures, such as changing the smoothing penalty function, the grid used to represent the PSF/PRF, the terms in the PSF/PRF shape parameter dependence, and many other tunable parameters. Our final analysis determined that not accounting for the PSF —i.e., approximating the illumination over each pixel as uniform— yielded the best photometry (highest precision). This could be because the PSF/PRF was found to be highly variable from one image to the next. One possible explanation for this lack of stability could be poor tracking. The Sony-$\alpha$ camera took thirty-second exposures but was attached to a HATNet mount, which was taking three-minute exposures. The dithering pattern (deliberately changing the telescope's pointing to smear the PSF) was designed for the HATNet mount. While this dithering pattern worked well over three minutes, it may not have been effective over thirty seconds. This could have caused PSF/PRF drift in every image, making modeling unpredictable. If proper tracking had been performed, we believe higher precision photometry could have been achieved using the correct PSF/PRF model.

\subsection{Future Endeavors}
\label{s:future}

\subsubsection{Browser-user-interface}
To ensure that citizen scientists will use this automated software suite, we are currently building a browser-user interface (BUI). This will allow citizen scientists to add their images via a locally hosted web browser and database and fully process them using this pipeline, automatically producing detrended light curves, as described in Section \ref{c:methodology}. Furthermore, we will supply diagnostic and visual tools to tune their configurations parameters via the BUI. The BUI will be implemented using Django, a Python-based web framework tool \footnote{\url{https://www.djangoproject.com/}}, and a variant of an SQL database (e.g., MySQL, SQLite).

\subsubsection{More features and functionalities}
This project is an ongoing software suite that will continuously receive updates and new features. Some features we plan to implement include (but are not limited to) the ability to import any raw image formats and convert them to FITS images, improving background detection, automatic target selections and light curve plotting, etc.

\section*{Acknowledgements}
This work was supported by the National Science Foundation under Grant No.
2311527.

Based on observations of the Hungarian-made Automated Telescope Network
and observations obtained at the Fred Lawrence Whipple Observatory of the
Smithsonian Astrophysical Observatory. 

This research has made use of the NASA
Exoplanet Archive, which is operated by the California Institute of Technology,
under contract with the National Aeronautics and Space Administration under the
Exoplanet Exploration Program.

This work has made use of data from the European Space Agency (ESA) mission
{\it Gaia} (\url{https://www.cosmos.esa.int/gaia}), processed by the {\it Gaia}
Data Processing and Analysis Consortium (DPAC,
\url{https://www.cosmos.esa.int/web/gaia/dpac/consortium}). Funding for the DPAC
has been provided by national institutions, in particular the institutions
participating in the {\it Gaia} Multilateral Agreement.

\vspace{5mm} \facilities{
    Gaia \citep{GAIA_2016, GAIA_2023}
%
%
}


\software{
    astropy \citep{astropy_13, astropy_18, astropy_22},
    fitsh \citep{Pal_12},
    astrometry.net \citep{Lang_et_al_10},
    AstroWISP \citep{Penev_et_al_25}
}

\bibliography{bibliography}{}
\bibliographystyle{aasjournal}



\end{document}